\documentclass[a4paper,11pt]{article}
\pdfoutput=1 
\usepackage{paper_jheppub}
\usepackage{subfigure}
\usepackage{tabularx}
\usepackage[T1]{fontenc}

\newcommand{\pT}{\ensuremath{p_{\mathrm{T}}}}
\newcommand{\mT}{\ensuremath{m_{\mathrm{T}}}}
\newcommand{\ET}{\ensuremath{E_{\mathrm{T}}}}
\newcommand{\MET}{\mbox{\ensuremath{\not \!\! \ET}}}
\newcommand{\PET}{\ensuremath{\vec p^{\textrm{miss}}_{\textrm{T}}}}
\def\ifb{\mbox{fb$^{-1}$}}
\def\ipb{\mbox{pb$^{-1}$}}
\def\GeV{\ifmmode {\mathrm{\ Ge\kern -0.1em V}}\else \textrm{Ge\kern -0.1em V}\fi}%

\title{\boldmath Opportunities and Challenges of Standard Model Production Cross Section Measurements in Proton--Proton Collisions at $\sqrt{s}$=8 TeV using CMS Open Data}

\author[a]{Aram Apyan}
\author[b]{William Cuozzo}
\author[b]{Markus Klute}
\author[b]{Yoshihiro Saito}
\author[b,c]{Matthias Schott\footnote{corresponding author}}
\author[b]{Bereket Sintayehu}

\affiliation[a]{Fermilab, USA}
\affiliation[b]{Massachusetts Institute of Technology, Cambridge, USA}
\affiliation[c]{Johannes Gutenberg-University, Mainz, Germany}

\emailAdd{matthias.schott@cern.ch}

\abstract{
The CMS Open Data project offers new opportunities to measure cross sections of standard model (SM) processes which have not been probed so far. We evaluate the challenges and the opportunities of the CMS Open Data project in the view of cross section measurements. In particular, we reevaluate the SM cross sections of the production of $W$ bosons, $Z$ bosons, top-quark pairs and $WZ$ dibosons in several decay channels at a center of mass energy of 8~TeV with an integrated luminosity of 1.8~fb$^{-1}$. These cross sections were previously measured by the ATLAS and CMS Collaborations and are used to validate our analysis and calibration strategy. The results indicate the achievable level of precision for future measurements using the CMS Open Data performed by scientists who are not members of the LHC Collaborations and hence lack detailed knowledge of experimental and detector related effects and their handling.
}


\begin{document}
\maketitle
\flushbottom

\section{\label{Sec:Intro}Introduction}

Precision measurements of standard model (SM) processes at the Large Hadron Collider (LHC) made tremendous progress in recent years. The differential measurement of the production cross sections of $W$ and $Z$ bosons, as well as top-quark pairs, reached a precision of a few percent (e.g. \cite{Aaboud:2016btc, Chatrchyan:2014mua, Chatrchyan:2013faa}), sometimes a few per-mil (e.g. \cite{Aad:2016izn, Aad:2015auj}). These form the basis for testing and improving modern Monte Carlo (MC) event generators that aim to describe those processes in high-energy hadron collisions. Numerous of these high-precision measurements are at the core of the research program of the ATLAS and CMS Collaborations since the beginning of the LHC. 

The CMS Collaboration has published significant amounts of recorded and simulated proton-proton collisions at a center-of-mass energy of 8~TeV within the context of the Open Data initiative \cite{CMS:OpenData}. In principle, the availability of these data sets allows physicists who are not member of the LHC Collaborations to perform measurements. With this work, we systematically evaluate the physics potential of the available CMS Open Data for cross section measurements and broaden the perspective of previous studies using CMS Open Data \cite{Cesarotti:2019nax, Tripathee:2017ybi}. Special focus is drawn on the limitations of such measurements and possible future improvements.

As a starting point, we estimate and derive several object calibration constants, either from previous publications, or using the data itself. In a second step, we measure several SM cross sections at a center-of-mass energy of 8~TeV with an integrated luminosity of 1.8\,\ifb and compare them to the results published by the CMS and ATLAS Collaborations. The agreement with the published results, as well as the assigned uncertainties on our measurements, indicate to which precision new cross section measurements can also be performed using CMS Open Data. 

The paper is structured as follows: The CMS detector, its physics objects, and the data-sets used in this analysis are summarized in Section \ref{Sec:CMS}. The calibration of the physics objects, such as electrons, muons or particle jets, is discussed in Section \ref{Sec:Calibration}. The cross section measurements of SM processes are discussed in Section \ref{Sec:BasicMeasurements}, leading to a discussion of the opportunities and challenges of cross section measurements within the CMS Open Data Initiative in Section \ref{Sec:Conclusion}.


\section{\label{Sec:CMS}The CMS Detector and CMS Open Data}

\subsection{The CMS Detector and Reconstructed Objects}

The data used in this analysis has been recorded with the CMS detector at the LHC in the year 2012. CMS is a typical high-energy physics experiment, using a superconducting solenoid of 6 m internal diameter with a magnetic field of 3.8 T~\cite{Chatrchyan:2008aa}. The inner detector (ID) of CMS can reconstruct trajectories of charged particles using silicon pixel and strip trackers. Electrons and photons are identified and measured in a crystal electromagnetic calorimeter (ECAL), while energies of hadrons or hadronic particle jets are determined in a brass/scintillator hadron calorimeter (HCAL). Muons are identified and measured in the muon system (MS), based on gaseous detectors, which surround the hadronic calorimeter and are embedded in the steel flux-return yoke of the magnet system. CMS uses a right-handed coordinate system. Its origin is defined at the interaction point of the proton collisions, the $x$ axis is pointing towards the center of the LHC, the $y$ axis pointing upwards and the $Z$ axis along the counterclockwise-beam direction. The polar angle $\theta$ is measured from the positive $z$ axis, however, mostly expressed in terms of the pseudorapidity $\eta$, defined by $\eta = - \ln(\tan\theta/2)$. The azimuthal angle $\phi$ is measured in the $x-y$ plane. We refer to \cite{Chatrchyan:2008aa} for a detailed description of the CMS experiment. CMS employs a particle-flow algorithm that provides a complete description of the event and identifies electrons, muons, photons, charged hadrons, and neutral hadrons \cite{CMS-PAS-PFT-09-001}.  


Electrons are identified as reconstructed energy clusters in the ECAL, which have been matched to tracks measured in the ID \cite{Adam:815410, Chatrchyan:2013dga, Khachatryan:2015hwa}. In this analysis, we typically require the transverse energy of electrons to be $\ET>25$ GeV within $|\eta|<1.44$ (barrel) or $1.57<|\eta|<2.5$ (endcap); the gap between barrel and endcap is determined by the detector layout. In addition, standard electron identification requirements, e.g. on the energy ratio measured in the ECAL and the HCAL or on the track impact parameters, are applied, following previous CMS measurements \cite{Chatrchyan:2014mua}. An electron passes a loose/tight isolation requirement if the vectorial sum of momenta of all reconstructed charged particles, stemming from the primary vertex, within a cone-size of $\Delta R=\sqrt{(\Delta \eta)^2+(\Delta \phi)^2}<0.3$ normalized by the $\ET$ of the electron is smaller than 0.15 (0.10) in the barrel (endcap).

Muons are reconstructed from a global fit of hits in the MS and the ID, seeded by tracks in the muon system \cite{CMS-PAS-MUO-10-002}. In this analysis, we typically require each muon to have a transverse momentum of $\pT>25$ GeV within a pseudorapidity range of $|\eta|<2.1$, corresponding to the single muon trigger coverage. In addition, following previous CMS measurements \cite{Chatrchyan:2014mua}, standard quality requirements on the numbers of hits in the ID and in the MS, on the $\chi^2$ of the fit, and on the track impact parameters are applied. A relative isolation variable is computed as described for the electrons for a loose and tight isolation definition, respectively, however, computed with a cone-radius of $\Delta R=0.4$.

Hadronic jets are reconstructed using an anti-k$_{\textrm{T}}$ algorithm with a radius parameter of 0.5 based on particle-flow objects \cite{CMS-PAS-PFT-10-001, CMS-PAS-PFT-09-001}, where the clustering algorithm rejects objects that are coming from vertices of additional interactions per bunch crossing (pile-up). A jet area method \cite{Chatrchyan:2011ds} is used to correct for the remaining pile-up contributions. 

In this analysis, we focus our study on jets with a transverse momentum of $\pT>30$ GeV and a rapidity of $|y|<2.4$, since this region allows for a good jet resolution and pile-up rejection. In addition, certain quality criteria on the reconstructed jet properties, such as energy fraction in the ECAL and HCAL or the number of particle-flow objects, are applied following standard CMS recommendations. Moreover, jets are required to have a distance of $\Delta R>0.5$ to all reconstructed electron, muon, and photon candidates. The three jets with the largest reconstructed $\pT$ are denoted as $j_1$, $j_2$, and $j_3$ in order of decreasing $\pT$ values. The origin of a jet from a bottom quark is identified via a combined secondary-vertex algorithm, which uses track impact parameter and secondary-vertex information \cite{Chatrchyan:2012jua}. In this analysis, we use a 'medium' working-point for identified b-jets with an average efficiency of 85\%.

Neutrinos leave the CMS detector undetected and hence cause an imbalance in the vectorial momentum sum of all final-state particles in the plane transverse to the beam axis. CMS defines the missing transverse energy as the negative vector sum of all $\pT$ of reconstructed particle flow objects, i.e. $\PET~=-\sum _{\textrm{PF}} \vec p_{\textrm{T}}$. The magnitude of \PET~ is denoted as \MET.

It is additionally required that selected reconstructed objects are not considered for further analysis if they are close to other reconstructed objects. Electron candidates are not further considered if a muon candidate with $\pT>20$ GeV, passing standard quality criteria, has been reconstructed within $\Delta R<0.3$ of the electron candidate. Jets are not further considered if a reconstructed muon or electron candidate with $\pT>20$ GeV and passing the above-mentioned quality criteria, are found within $\Delta R<0.3$ of the jet candidate.


\subsection{Software and Infrastructure}

The CMS Open Data Software Framework (Release CMSSW\_5\_3\_32), available in \cite{CMS:OpenData}, builds the basis of this analysis. A dedicated open-source framework, \textsc{Bacon} \cite{cit:bacon}, which was used for several published studies of the CMS Collaboration, e.g. \cite{Chatrchyan:2014mua}, is used to read the Analysis Object Data (AOD) \cite{Hinzmann:2011zz}, extracting information on reconstructed objects as well as particle-level data\footnote{Particle-level information denotes the available information on final states after the MC event generation step, i.e. prior to the detector simulation}, if available. The \textsc{Bacon} software framework is also used to apply a selection of events which have been recorded under stable detector conditions, known as \textit{GoodRun-List} \cite{CMS:GoodRunList}, provided by the CMS Open Data project, leading to a reduced output-format based on the \textsc{Root} software framework \cite{Brun:1997pa}. The typical size of one simulated top-quark pair event in the \textsc{Bacon}-output format amounts to 5 kB. For this work, we developed an additional software package, which reduces the output files of \textsc{Bacon} further and transforms them into a plain \textsc{Root}-NTuple, denoted as \textsc{ODNTuple} in the following with an average event size of 0.8 kB. Our analysis is based on these \textsc{ODNTuple} data. 

\subsection{\label{sec:SelOD}Selected Open Data}

The data acquisition system of CMS records only the event information of collisions with dedicated signatures due to the high-collision rate and the limited bandwidth for data-processing. The data used in this analysis has been collected when one of the triggers \textit{HLT\_IsoMu24}, \textit{IsoMu24\_eta2p1} or \textit{HLT\_Ele27\_WP80} has fired. These triggers are unprescaled for the full 2012 data-set and aim to collect events with at least one isolated muon candidate within $|\eta|<2.4$ and $\pT>24$ GeV or with at least one electron candidate within $|\eta|<2.5$ and $\ET>27$ GeV.

In total, muon-triggered (electron-triggered) events corresponding to 1.83\,\ifb (1.78\,\ifb) \cite{CMS:MuData1, CMS:MuData2} of integrated luminosity from CMS Open Data \cite{CMS:EData1, CMS:EData2} have been processed (Table \ref{tab:DataSamples}). We only studied roughly 10\% of the full available dataset due to limitations on the available computing resources during this project as well as the fact that our final results are already dominated by systematic uncertainties. The integrated luminosity has been calculated using the publicly available \textit{GoodRun-List}. We assume an uncertainty of 2.5\% in the integrated luminosity following the official CMS recommendation (\cite{CMS:2013gfa}). 

\begin{table}[tb]
\footnotesize
\begin{center}
\begin{tabular}
{l | l | c}
\hline
Data stream / trigger name							& Dataset Name										&	$\int L dt $ [\ipb]	\\
\hline
single muon trigger									& /SingleMu/Run2012C-22Jan2013-v1 \cite{CMS:MuData1}		&	1,828			\\
(HLT\_IsoMu24,									& /SingleMu/Run2012B-22Jan2013-v1 \cite{CMS:MuData2}		&					\\
HLT\_IsoMu24\_eta2p1)								&													&					\\
\hline
single electron trigger 								& /SingleElectron/Run2012B-22Jan2013-v1 \cite{CMS:EData1}		&	1,776			\\
(HLT\_Ele27\_WP80)								& /SingleElectron/Run2012C-22Jan2013-v1 \cite{CMS:EData2}		&					\\
\hline
\hline
\end{tabular}
\caption{Overview of data samples used in this analysis together with the corresponding integrated luminosity and the triggers, which have been used during the data taking.\label{tab:DataSamples}}
\end{center}
\end{table}

\subsection{MC Simulated Samples}

An overview of the various signal and background samples used in this analysis is given in Table \ref{tab:MCSamples}, indicating the underlying physics process, the dataset name, and the corresponding inclusive cross section at next-to-leading order (NLO) or next-to-next-to-leading order (NNLO). The Drell--Yan processes ($W/Z$) in the electron and muon decay channels were generated with the \textsc{PowhegBox} v.1.0 MC program~\cite{Alioli:2008gx, Alioli:2010xd} interfaced to the \textsc{Pythia} v.6.4.26 parton shower model~\cite{Sjostrand:2006za}. All other processes are modeled with the tree-level matrix element event generator \textsc{MadGraph} v5.1.3.30 \cite{Alwall:2011uj} interfaced with PYTHIA 6.4.26. The CT10 parton distribution functions (PDFS) \cite{Gao:2013xoa} and the Z2* \textsc{Pythia}6 tune~\cite{Chatrchyan:2013gfi,Khachatryan:2015pea} are used. The decays of tau-leptons are modeled using the \textsc{Tauola} program \cite{Davidson:2010rw}. The \textsc{Pythia} v.6.4.26 is used for the modeling of photon radiation off final state particles. The strong coupling constant $\alpha_s$ has been set to $0.130$ at the $Z$ boson mass scale for all matrix element calculations. The effect of pile-up has been simulated by overlaying MC-generated minimum bias events. The \textsc{Geant4} program was used to simulate the passage of particles through the CMS detector \cite{Agostinelli:2002hh}. 

The simulated event samples are reweighted to describe the distribution of the number of pile-up events in the data by reweighting the $\rho$ parameter distribution, where $\rho$ denotes the diffuse offset energy density~\cite{Khachatryan:2016kdb}. Moreover, a reweighting of the longitudinal position of the primary $pp$ collision vertex of the MC samples to data has been performed. The resulting $\rho$ distributions for data and simulated $Z$ boson samples in the electron and muon decay channels are shown in Figure \ref{fig:PileUp}. The differences in MC predictions with and without reweighting is taken as a systematic uncertainty. 

\begin{table}[tb]
\footnotesize
\begin{center}
\begin{tabularx}{\textwidth}{l | l | c | c}
\hline
Process											& Dataset Name													&	Inclusive		& order \\
												&																&	$\sigma$ [pb]	& in $\alpha_s$	\\
\hline
$pp \rightarrow Z/\gamma^*+X \rightarrow e^+e^- + X$			& DYToEE\_M-20\_CT10\_TuneZ2star\_v2\_8TeV \cite{CMS:MCZee}		&	1916		& NNLO			\\
$pp \rightarrow Z/\gamma^*+X \rightarrow \mu^+\mu^- + X$		& DYToMuMu\_M-20\_CT10\_TuneZ2star\_v2\_8TeV \cite{CMS:MCZmumu}	&	1916		& NNLO			\\
$pp \rightarrow Z/\gamma^*+X \rightarrow l^+l^- + X$			& DYJetsToLL\_M-50\_TuneZ2Star\_8TeV	\cite{CMS:MCZLL}				& 	3533		& NNLO			\\
\hline
$pp \rightarrow W^++X \rightarrow \mu^+\nu + X$				& WplusToMuNu\_CT10\_8TeV	\cite{CMS:MCWPMunu}					&	7322		& NNLO			\\
$pp \rightarrow W^-+X \rightarrow \mu^-\nu + X$				& WminusToMuNu\_CT10\_8TeV \cite{CMS:MCWMMunu}				&	5181		& NNLO			\\
$pp \rightarrow W^++X \rightarrow \tau^+\nu + X$				& WplusToTauNu\_CT10\_8TeV \cite{CMS:MCWPTau}					&	7322		& NNLO			\\
$pp \rightarrow W^-+X \rightarrow \tau^-\nu + X$				& WminusToTauNu\_CT10\_8TeV \cite{CMS:MCWMTau}					&	5181		& NNLO			\\
\hline
$pp \rightarrow t\bar{t}+X \rightarrow 2l2\nu2b+ X$				& TTJets\_FullLeptMGDecays\_TuneP11TeV\_8TeV	\cite{CMS:MCTTLep}	&	112.3	& NLO			\\
$pp \rightarrow t\bar{t}+X \rightarrow 1l1\nu2q2b+ X$			& TTJets\_SemiLeptMGDecays\_8TeV \cite{CMS:MCTTSemi}				&	107.2	& NLO			\\
$pp \rightarrow t\bar{t}+X \rightarrow 4q2b+ X$					& TTJets\_HadronicMGDecays\_TuneP11mpiHi\_8TeV \cite{CMS:MCTTHad}	&	25.8		& NLO			\\
\hline
$pp \rightarrow WW+X \rightarrow 2l2\nu+ X$					& WWJetsTo2L2Nu\_TuneZ2star\_8TeV \cite{CMS:MCWW}				&	5.8		& NLO			\\
$pp \rightarrow WZ+X \rightarrow 3l1\nu+ X$					& WZJetsTo3LNu\_8TeV\_TuneZ2Star \cite{CMS:MCWZ}					&	1.1		& NNLO			\\
$pp \rightarrow ZZ+X \rightarrow 4\mu+ X$					& ZZTo4mu\_8TeV \cite{CMS:MCZZ}									&	0.077	& NLO			\\
\hline
\end{tabularx}
\caption{Overview of simulated event samples used in this analysis together with the corresponding inclusive cross sections. Inclusive charged leptons ($e,\mu,\tau$) are denoted with $l$.\label{tab:MCSamples}}
\end{center}
\end{table}

\begin{figure}[tb]
\begin{center}
\includegraphics[width=7.3cm]{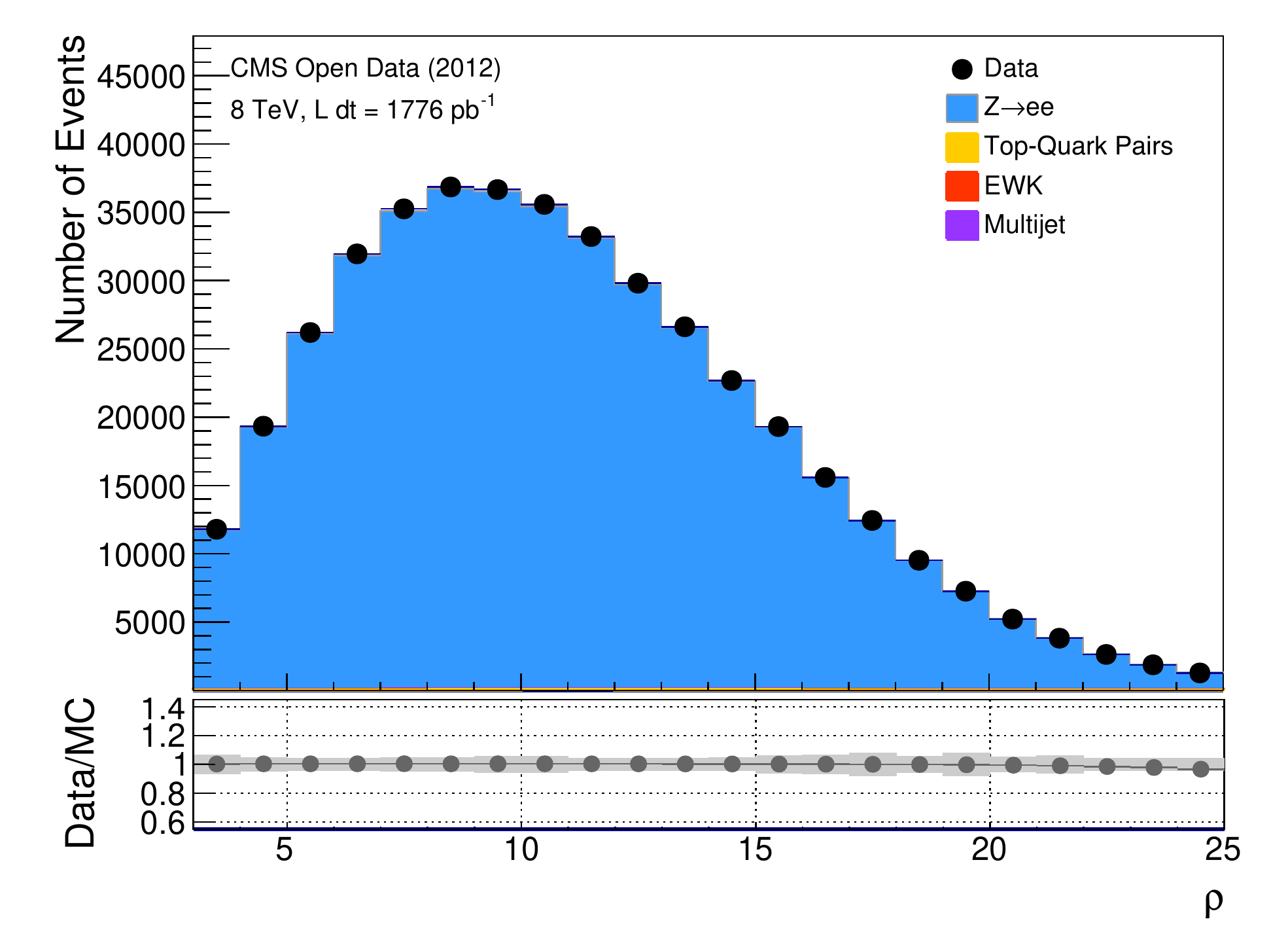}
\hspace{0.1cm}
\includegraphics[width=7.3cm]{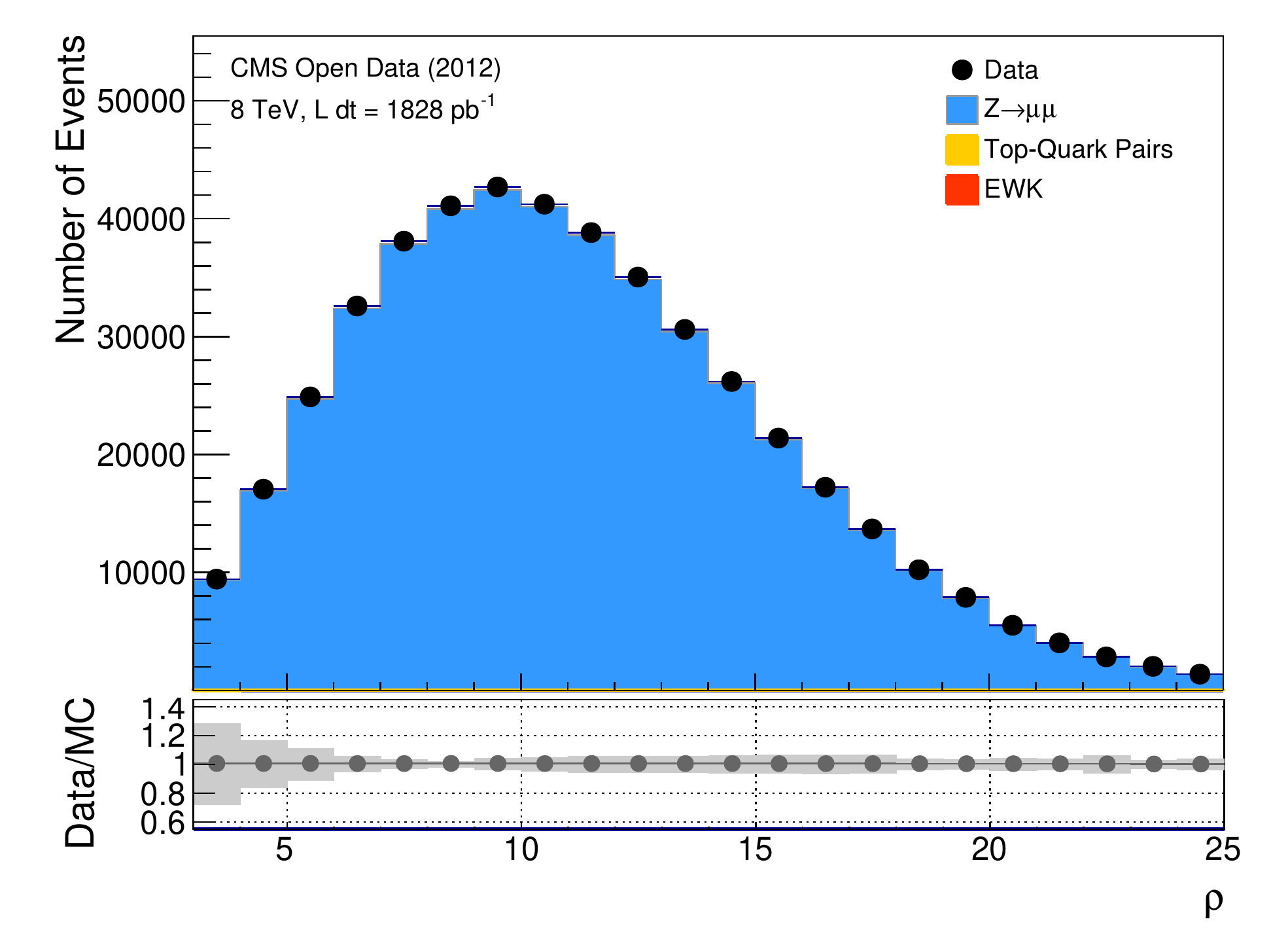} 
\caption{\label{fig:PileUp} Distribution of the $\rho$ parameter, sensitive to the pile-up activity, per event for electron (left) and muon (right) events as well as reweighted simulated Drell--Yan events in electron and muon decay channels, respectively. The gray band in the ratio indicates the systematic uncertainties on the predictions without luminosity uncertainties.}
\end{center}
\end{figure}


\section{\label{Sec:Calibration}Calibration}

Although the detector simulation of CMS experiments provides a very good description of the expected event signatures, some remaining differences in reconstruction, trigger, and isolation efficiencies, as well as in the momentum and energy scales and resolutions are present between MC and data. Dedicated corrections are applied to minimize these differences and are discussed in the following. The $Z$ boson candidate events in the electron and muon decay channels in data and simulated samples are used to validate our corrections. The $Z\rightarrow \mu^+\mu^-$ candidate events are selected by requiring events with exactly two oppositely charged, isolated muons with $\pT>25$ GeV and $|\eta|<2.1$. The $Z\rightarrow e^+e^-$ candidate events are selected by requiring events with exactly two oppositely charged, isolated electrons with $\ET>$ GeV and $|\eta|<1.44$ or $1.57<|\eta|<2.5$. The invariant mass of these two lepton candidates has to be between 60 and 120 GeV. This selection ensures a nearly background-free selection of $Z$ boson candidates. 

\subsection{Muon Performance}

The momentum scale and resolution corrections for muons are derived by comparing the reconstructed invariant mass spectrum of $Z$ boson candidates between data and simulation. The transverse momenta of the reconstructed muons can be modified via 

\begin{equation}
p^{\prime \mathrm{Reco}}_{\textrm{T}} = p^{\mathrm{Truth}}_{\textrm{T}} + \beta \cdot (\alpha \cdot p^{\mathrm{Reco}}_{\textrm{T}}-p^{\mathrm{Truth}}_{\textrm{T}}),
\end{equation}

\noindent where $p^{Reco}_{\textrm{T}}$ is the reconstructed muon momentum, $p^{Truth}_{\textrm{T}}$ is the truth muon momentum on particle level, $\alpha$ is a momentum scale parameter, and $\beta$ is a resolution parameter. The parameters $\alpha$ and $\beta$ are determined for three different regions in $\eta$, corresponding to the two endcap and one barrel regions, by a $\chi^2$ minimization procedure. The $\chi^2$ is calculated between the invariant mass spectrum of the di-lepton system in $Z$ boson events in data and simulation for different choices of $\alpha$ and $\beta$. The average values of $\alpha$ and $\beta$ were found to be 0.998 and 1.13, respectively. Uncertainties in the momentum scale of 0.002 in the barrel region and 0.003 in the endcap regions are applied, which cover the observed discrepancies with data. The uncertainty in the resolution parameter is 0.05. The size of the uncertainties have been additionally tested by varying the invariant mass window requirement. The comparison between data and MC of the invariant mass distribution of di-muon pairs after the calibration procedure is shown in Figure \ref{fig:MuonCalib}.

The corrections for reconstruction and trigger efficiencies for single muons as well as their uncertainties were taken from official CMS publications \cite{Khachatryan:2016bia, Chatrchyan:2014mua}. The average correction weights of the muon reconstruction and trigger efficiencies are found to be $0.985\pm0.006$ and $0.950\pm0.008$, respectively. The muon isolation is well described by the simulation, i.e. the correction weight is set to $1.000$ and an uncertainty of 0.002 is applied. A comparison of the $\eta$ distribution of muons from $Z$ boson candidates between data and MC is shown in Figure \ref{fig:MuonCalib}, where all the corrections above have been applied. The remaining differences are covered by the systematic uncertainties.

\begin{figure}[tb]
\begin{center}
\includegraphics[width=7.3cm]{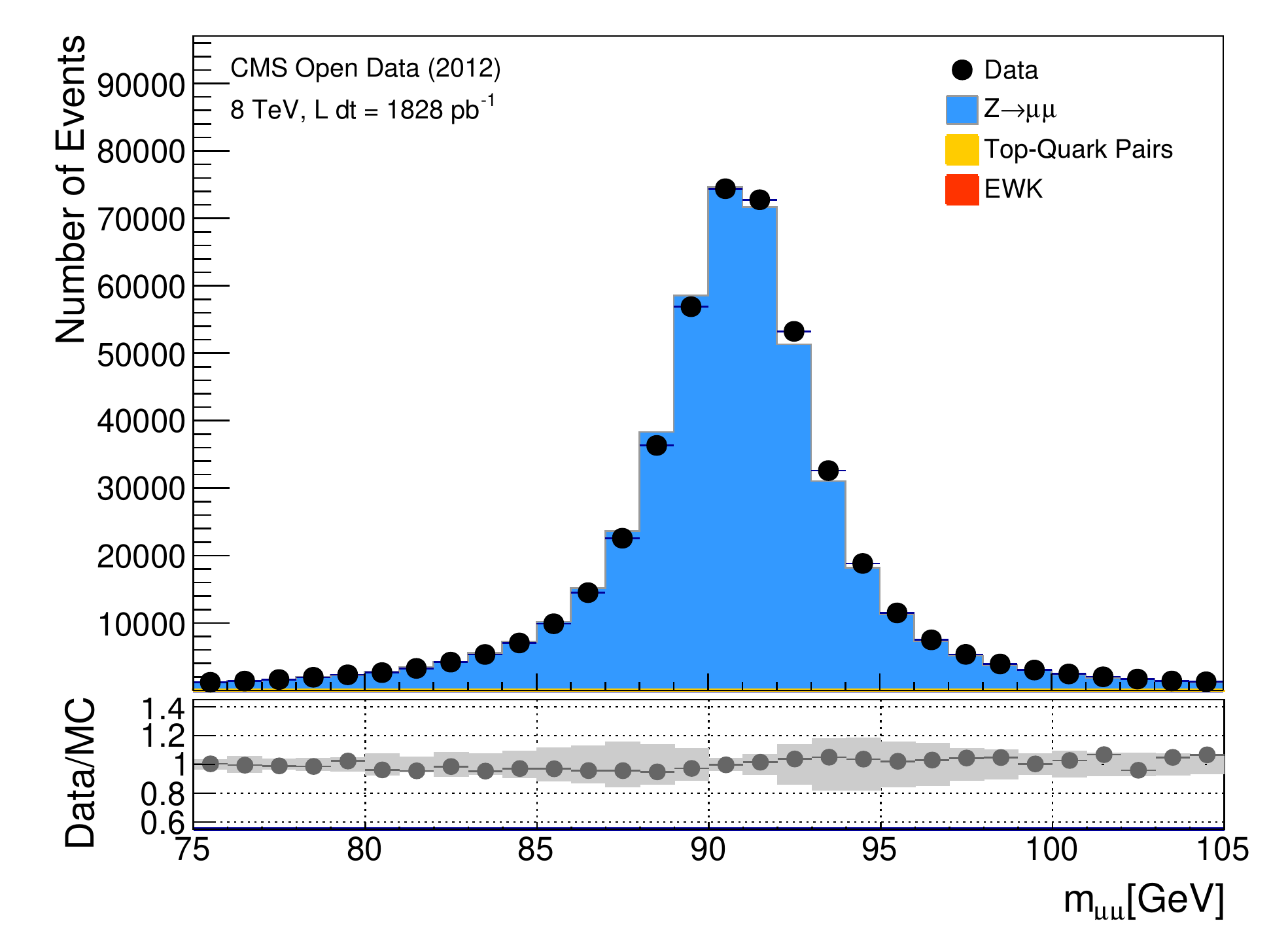} 
\hspace{0.1cm}
\includegraphics[width=7.3cm]{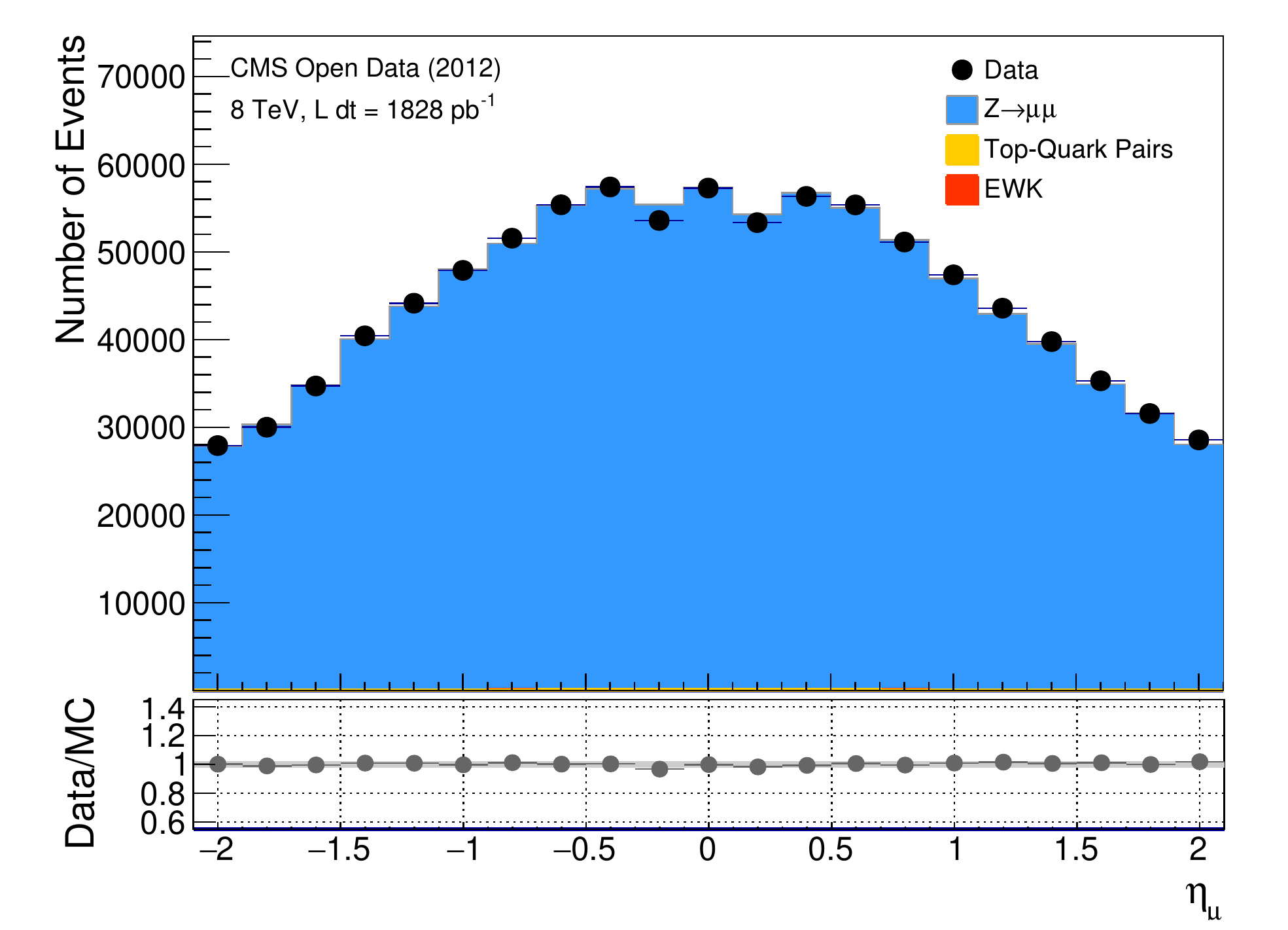}
\caption{\label{fig:MuonCalib} Comparison of the di-muon mass spectrum (left) and the muon $\eta$ distribution (right) for data and MC for $Z$ boson candidate events, after all corrections have been applied. The gray band in the ratio indicates the systematic uncertainties on the predictions without luminosity uncertainties.}
\end{center}
\end{figure}

\subsection{Electron Performance}

The energy scale and resolution corrections for electrons are derived in a similar way as for the muons, however, an off-set parameter $\kappa$ is also used in addition to a multiplicative scale factor $\alpha$. The addition of the $\kappa$ off-set parameter leads to a significantly better description of the data with the MC simulated samples. Using
\begin{equation}
E^{\prime \mathrm{Reco}}_{\textrm{T}} = \kappa + E^{\mathrm{Truth}}_{\textrm{T}} + \beta \cdot (\alpha \cdot E^{\mathrm{Reco}}_{\textrm{T}}-E^{\mathrm{Truth}}_{\textrm{T}}),
\end{equation}
we find an average value of the energy offset parameter of $\kappa =-0.5\pm0.1$ GeV and values of $\alpha$ and $\beta$ consistent with 1. The systematic uncertainty in the scale parameter $\alpha$ is 0.003, while the resolution uncertainties on $\beta$ range from 0.017 (barrel) to 0.045 (endcap) for electrons with $\ET<80$ GeV, and are 0.005 for $\ET>80$ GeV \cite{Khachatryan:2015hwa}.

Corrections to the electron identification and the single-electron trigger efficiencies are taken from \cite{Khachatryan:2015hwa, Chatrchyan:2014mua} and are close to 1 in most regions, with larger corrections for electrons with $\ET<30$ GeV in parts of the endcap region. The applied systematic uncertainties are 1.4\% and 2.4\% for electron identification and trigger efficiencies, respectively. The isolation for electron is also well described by the MC simulations, so no reweighting is necessary for the isolation requirement efficiency. Given the more complicated nature of electron signatures compared to muons in the ID and ECAL, an uncertainty of 0.004 in the isolation requirement efficiency is applied. A comparison between data and MC of the invariant mass distribution and of the $\eta$ distribution of electrons from $Z$ boson candidates is shown in Figure \ref{fig:ElectronCalib}, where all corrections have been applied. The remaining differences are covered by the systematic uncertainties.

\begin{figure}[tb]
\begin{center}
\includegraphics[width=7.3cm]{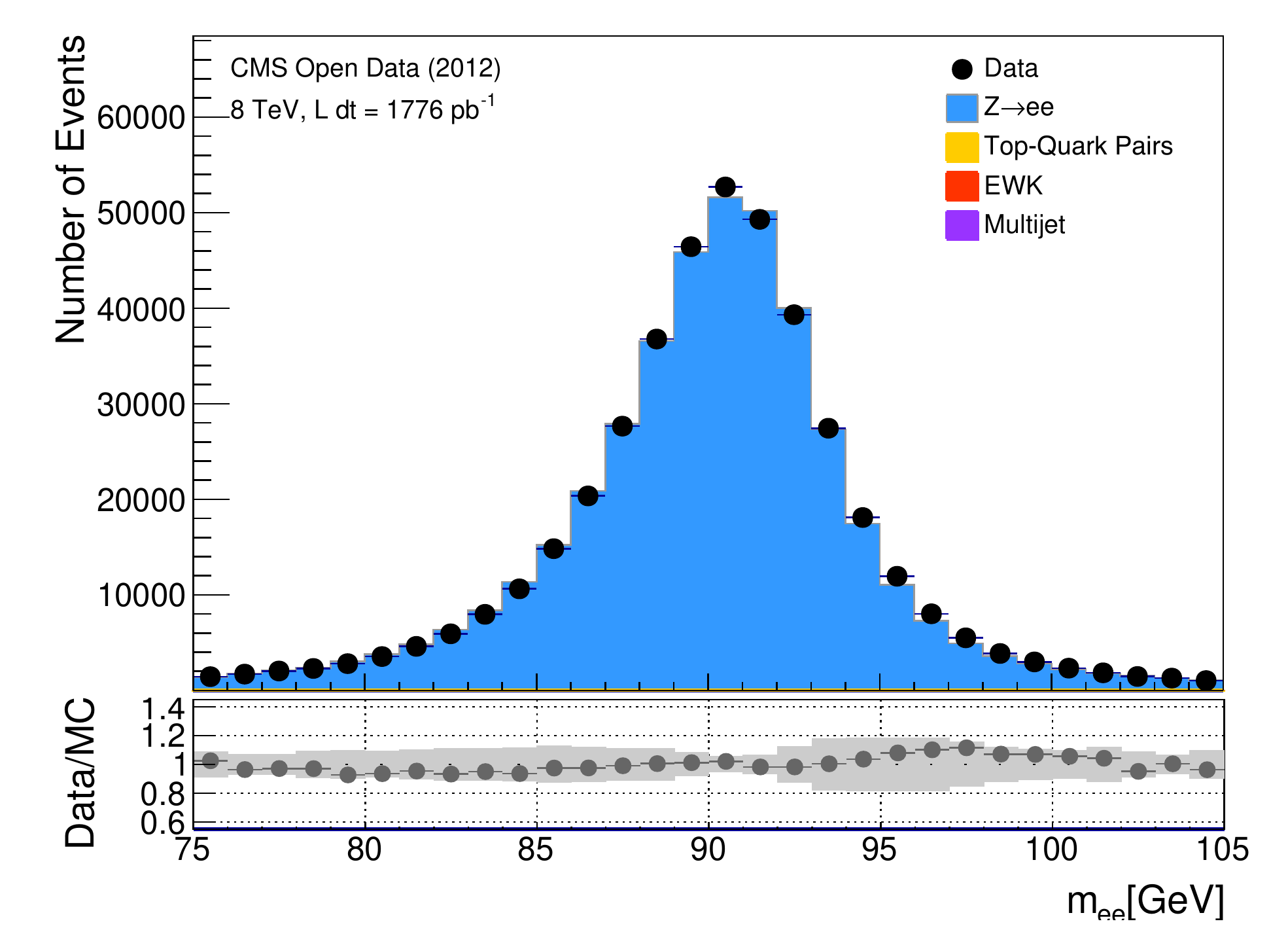} 
\hspace{0.1cm}
\includegraphics[width=7.3cm]{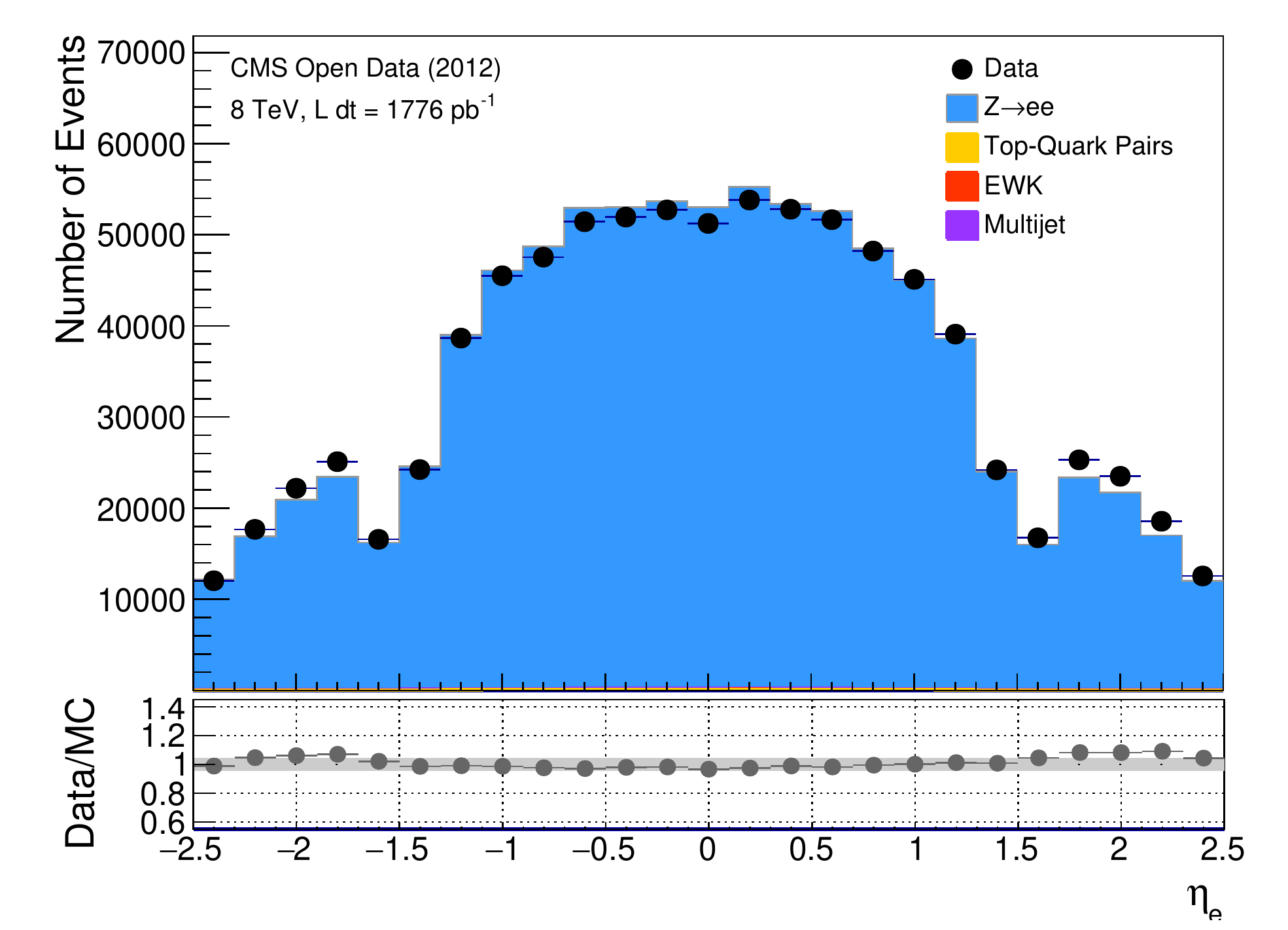}
\caption{\label{fig:ElectronCalib} Comparison of the di-electron mass spectrum (left) and the electron $\eta$ distribution (right) for data and MC for $Z$ boson candidate events, after all corrections have been applied. The gray band in the ratio indicates the systematic uncertainties on the predictions without luminosity uncertainties. }
\end{center}
\end{figure}

\subsection{Jet Energy Scale and Resolution}

The official CMS calibration and corrections for particle jets, in particular, the jet energy scale (JES) and the jet energy resolution (JER), have been applied within the \textsc{Bacon} framework. These jet corrections and uncertainties were derived from the simulation, and are confirmed with in situ measurements using the energy balance of dijet and photon+jet events \cite{Khachatryan:2016kdb}. A reduced set of systematic variations is used to estimate JES and JER uncertainties on the final measurement. In particular, the JES is varied by 2\% for $|y^{jet}|<1.3$ and by 3\% for $|y^{jet}|>1.3$, following \cite{Khachatryan:2016kdb}. The JER is varied by 20\% for $30<\ET^{jet}<100$ GeV, by 10\% for $100<\ET^{jet}<1000$ GeV and by 5\% for $\ET^{jet}>1000$ GeV. We apply a JER uncertainty of 20\% for jets with an absolute rapidity larger than 2.1. This simplified treatment of the jet calibration does not allow for a correct evaluation of correlations between different phase space regions. However, the resulting systematic uncertainties in the inclusive cross section measurements are expected to be valid. 

The calibration of jets, as well as the assigned systematic uncertainties, is tested again using $Z$ boson candidate events in the muon decay channel. For this, $Z$ boson events with a transverse momentum, $\pT(Z)$, between 50 and 100 GeV with exactly one reconstructed jet with $\ET^{jet}>30$ GeV and $|y^{jet}|<2.1$ are selected. The transverse momentum of the $Z$ boson, precisely measured by its decay leptons, should be balanced in a first approximation by the transverse energy of this jet, hence the ratio of $\pT(Z)/\ET^{jet}$ should peak around 1. The comparison between data and MC of this ratio is shown in Fig. \ref{fig:JetPerformance}, where a good agreement within the assigned systematic uncertainties can be seen. This study has been repeated for higher values of $\pT(Z)$ and more jets in the final state, all indicating a good closure.

\subsection{Tagging of b-Quarks}

The b-tagging efficiency for the working point of the Combined Secondary Vertex algorithm used in this analysis is 70\% for a light-quark misidentification probability of 1.5\%. The efficiency has been measured in data and compared with the MC predictions \cite{Chatrchyan:2012jua, Khachatryan:2016mqs}. In general, a very good agreement has been found for jet energies between 30 and 500 GeV, where a systematic uncertainty of the order of 3\% was assigned on the efficiency estimate in data. Hence, we do not apply additional b-tagging efficiency corrections. However, we assign an uncertainty of 5\% in the b-tagging efficiency as we do not apply any kinematic dependent efficiency corrections.

\subsection{Missing Transverse Energy}

We apply the official CMS calibration constants and correction factors to the reconstructed \MET~observable on an event-by-event basis. The assigned uncertainties in \MET\, are based on \cite{Khachatryan:2014gga}, where the scale uncertainty is taken to be 10\% for $\MET<20$ GeV, 5\% for $20<\MET<100$ GeV and 2\% for $\MET>100$ GeV. The uncertainty in the \MET\, resolution is applied as a function of the scalar sum of all transverse energies of all reconstructed hadronic objects in the event, $\sum \ET$, and is taken to be 20\% for $\sum \ET<100$ GeV and 10\% for $\sum \ET<100$ GeV. In addition, we propagate all the jet energy scale and resolution uncertainties to \MET by studying the impact on a recalculated \MET~observable, which is based on all reconstructed objects in the event. 

The \MET observable is validated by studying the observed hadronic recoil, the vector sum of all hadronic energies, in events containing $Z\rightarrow\mu\mu$ candidates. By construction, the transverse momentum of the $Z$ boson must be balanced by the hadronic recoil, $\vec{HR}$, i.e. $\vec \pT(Z)+\vec{HR} = 0$. Since the momenta of the decay muons of the $Z$ boson can be measured with high precision, providing a high precision measurement of $\pT(Z)$, one can effectively probe the simulation of the detector response on the hadronic recoil. It should be noted that the description of the hadronic recoil is effectively the same as the description of \MET. This can be seen most easily in $W$ boson events in their leptonic decay channel. Momentum conservation in the transverse plane implies $\vec \pT(W)+\vec{HR} = \vec \pT^l + \vec \pT^\nu + \vec{HR} = 0$, hence $\vec \pT^\nu = - \vec \pT^l - \vec{HR}$. The \MET description is  therefore equivalent to the description of $|\vec HR|$ as the decay lepton can be measured with high precision. A comparison of the \MET~distribution in $Z$ boson events, defined as $|\vec \pT(Z)+\vec{HR}|$, in the muon decay channel with $\pT(Z)<30$ for data and MC is shown in Figure \ref{fig:METPerformance}, together with the systematic uncertainties. The observed differences are covered by the systematic uncertainties.

\begin{figure}[t]
\centering
\begin{minipage}{7.3cm}
  \centering
  \includegraphics[width=1.0\linewidth]{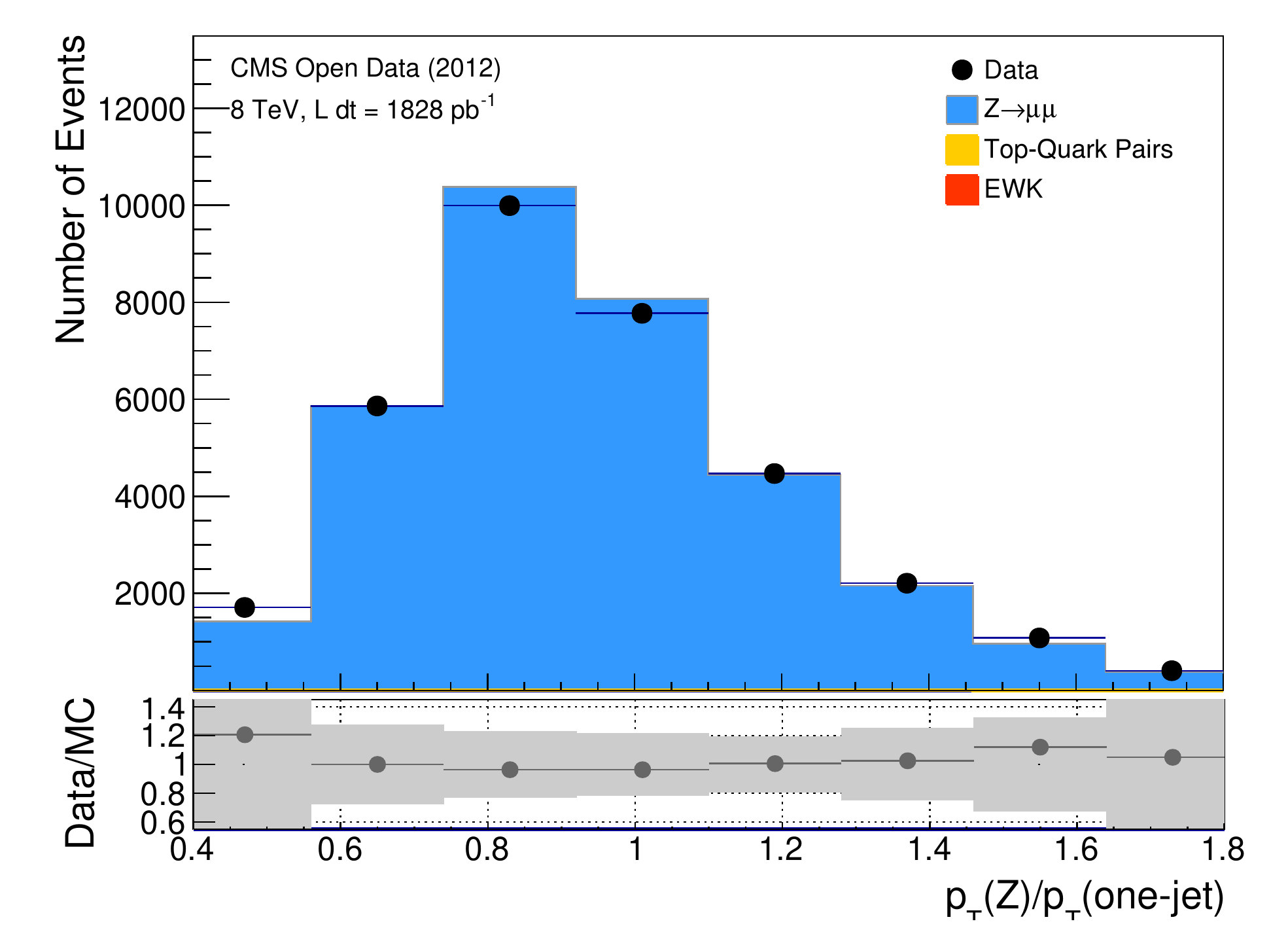}
	\caption{Comparison of the ratio of measured $\pT(Z)$ and the measured jet energy $\ET$ for $Z$ boson events in the muon decay channel with exactly one jet with $\ET>30$ GeV and $50<\pT(Z)<100$  GeV for data and MC, after all corrections have been applied. The gray band in the ratio indicates the systematic uncertainties on the predictions without luminosity uncertainties.\label{fig:JetPerformance}}
\end{minipage}%
\hspace{0.1cm}
\begin{minipage}{7.3cm}
\centering
\includegraphics[width=1.0\linewidth]{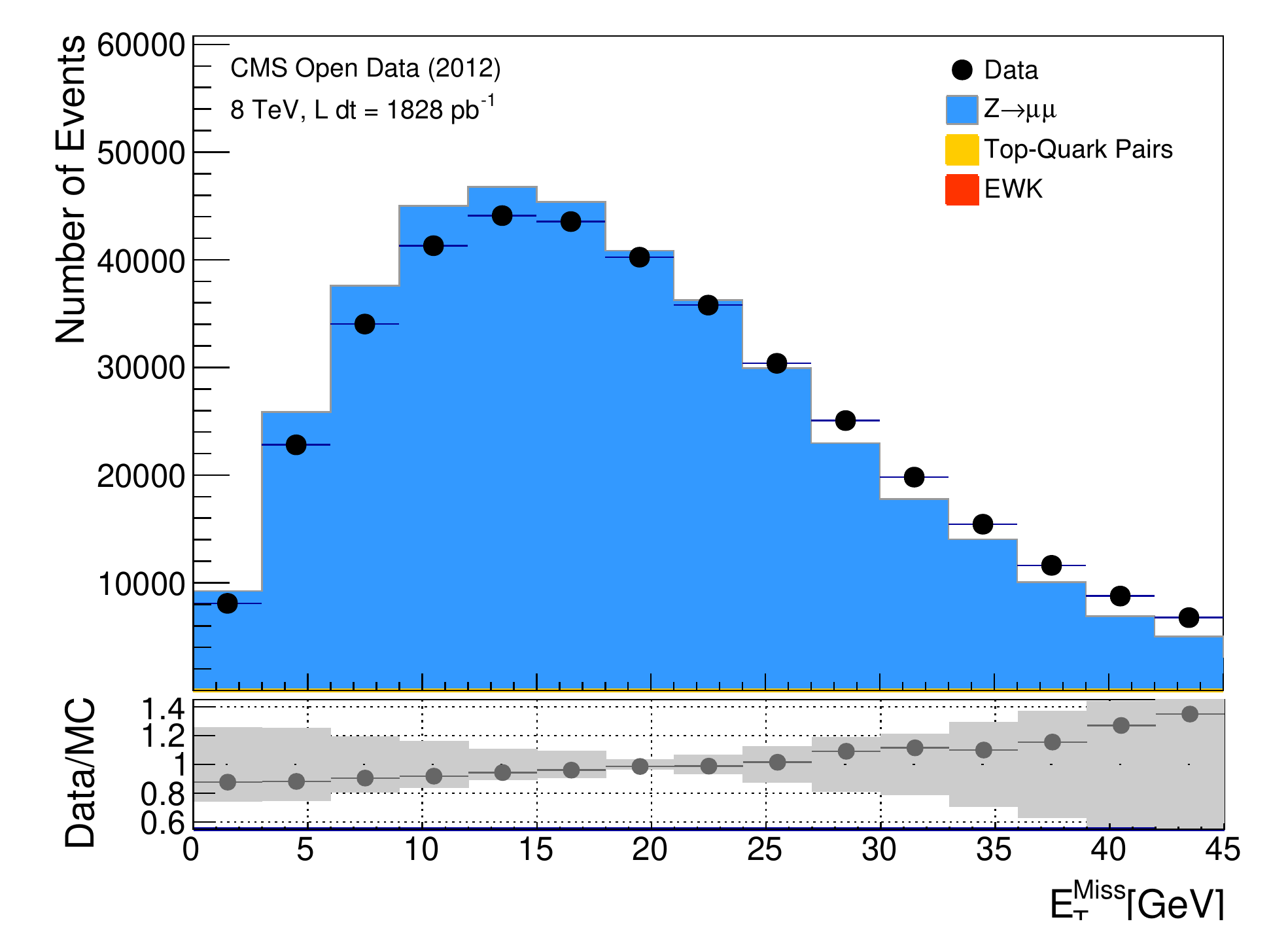}
\caption{Comparison of the \MET distribution for $Z$ boson events in the muon decay channel with $\pT(Z)<30$ GeV for data and MC, after all corrections have been applied. The gray band in the ratio indicates the systematic uncertainties on the predictions without luminosity uncertainties. \vspace{0.94cm}\label{fig:METPerformance}}
\end{minipage}
\end{figure}


\section{\label{Sec:BasicMeasurements}Standard Model Cross Section Measurements}

\subsection{Standard Processes and Signal Selection}

To validate all aspects of our analysis framework, starting from the correct interpretation of the reconstructed objects to the estimation of systematic uncertainties, several inclusive cross section measurements of SM processes have been performed and compared to high-precision measurements of the LHC Collaborations as well as to the theoretical predictions. In particular, we have measured the fiducial cross section of the Drell--Yan process in the electron and muon decay channel, $pp\rightarrow Z/\gamma^*\rightarrow l^+l^-$ ($l=e,\mu$), the fiducial cross section of $W^\pm$ boson production in the muon decay channels, $pp\rightarrow W^{\pm}\rightarrow \mu^\pm \nu$, the production cross section of the $Z$ boson in association with exactly one energetic jet, the production of top-quark pairs in the electron-muon final state, and the $WZ$ diboson production cross section in the fully leptonic decay channel. These processes probe different aspects of the analysis infrastructure. The $Z$ boson production cross section mainly probes lepton identification and reconstruction efficiencies, as well as the jet calibration when requiring an additional energetic jet in the final state. The study of $W$ bosons also probes the single lepton trigger performance. The top-quark pair production enables tests of the jet performance and the b-tagging performance. The study of the $WZ$ diboson production is mainly sensitive to the lepton reconstruction performance. 

The inclusive production cross section for a given process can be experimentally determined via
\begin{equation}
\sigma_{V}^{\textrm{incl.}} = \frac{N_{\textrm{Signal}}}{\epsilon \cdot BR \cdot \int L dt} = \frac{N_{\textrm{Signal}}}{A \cdot C \cdot BR \cdot \int L dt}\,.
\end{equation}
The number of signal events is given by $N_{Signal} = N_{Data}-N_{Bkg}$, where $N_{Data}$ is the number of selected events in data and $N_{Bkg}$ is the number of background events surviving the signal selection.
The factor $\epsilon$ is the efficiency of the signal events passing the signal selection criteria. The efficiency correction $\epsilon$ can further be decomposed as the product of a fiducial acceptance, $A$, and a detector-induced correction factor, $C$, i.e. $\epsilon = A \cdot C$, defined below. To correct the cross section for the choice of a specific decay channel, a branching ratio factor $BR$ is applied, which is known to high precision. Finally, the event yield is normalized by the integrated luminosity $\int L dt$ of the data sample analyzed.

The efficiency correction factor $\epsilon$ can be estimated with simulations of the signal process. These simulations include both a detailed description of the object reconstruction in the detector, called the \textit{detector level}, and the final-state particle information of the generator calculations, called the \textit{particle level}. The same signal selection cuts that are applied on data are applied to the simulated events at detector level. In addition, basic signal selection cuts, such as a minimum $\pT$ cut, can also be applied to the final-state particles at the particle level. Following these definitions, $\epsilon$ can be defined as the ratio of all events which pass the signal selection on detector level $N^{\textrm{selected}}_{\textrm{detector}}$ over the number of all generated events $N^{\textrm{all}}_{\textrm{particle}}$. 

The fiducial acceptance, $A$, and a detector-induced correction factor, $C$, are defined as $A=N^{\textrm{selected}}_{\textrm{particle level}}/N^{all}_{\textrm{particle level}}$ and $C= N^{\textrm{selected}}_{\textrm{detector level}}/N^{\textrm{selected}}_{\textrm{particle level}}$, respectively. The fiducial acceptance $A$ is therefore the ratio of the number of events that pass the geometrical and kinematic cuts of an analysis at particle level, $N^{\textrm{selected}}_{\textrm{particle level}}$, over the total number of generated events in a simulated sample of signal process, $N^{\textrm{all}}_{\textrm{particle level}}$. These selection cuts on particle level require geometrical and kinematic constraints close to the cuts applied on the reconstructed objects at detector level, e.g. muons in the final state should fulfill $\pT>25$ GeV and $|\eta | < 2.1$. The dominant uncertainties on the fiducial acceptance are renormalisation and factorisation scale uncertainties as well as uncertainties due to the limited knowledge of the parton distribution function (PDF) of the proton. The detector response correction factor $C$ is the ratio of selected events on detector level, $N^{\textrm{selected}}_{\textrm{detector level}}$, over all events that pass the kinematic selection on particle level, i.e. $N^{\textrm{selected}}_{\textrm{particle level}}$. The fiducial cross section for a given process can therefore be measured by

\begin{equation}
\label{EQN:CrossSectionExp}
\sigma_{V}^{\textrm{fid.}} = \frac{N_{\textrm{Signal}}}{C \cdot BR \cdot \int L dt}.
\end{equation}
Since this definition is independent of $A$, modelling uncertainties are minimized. Once the fiducial cross section is known, one can extrapolate to the inclusive cross section using the acceptance correction factor, i.e. $\sigma_{V}^{\textrm{incl.}} = \sigma_{V}^{\textrm{fid.}} / A$.

All recorded events that are kept for the analysis are required to fulfill the \textit{GoodRun-List} requirements, contain at least one good primary vertex, and to be either triggered by a single electron or a single muon trigger (see Section~\ref{sec:SelOD}).

The selection of the $Z$ boson candidates has already been introduced in Section \ref{Sec:Calibration}. The fiducial volume for the Drell-Yan process is defined at so called Born level (before final state photon radiation) by requiring the two decay leptons with a transverse momentum of $\pT>$ 25 GeV within $|\eta|<2.1$ and $|\eta|<2.4$ for the muon and electron decay channels, respectively, following the corresponding CMS publication \cite{Chatrchyan:2014mua}. When studying $Z$ boson production in association with jets in the muon decay channel, the minimum lepton $\pT$ requirement is lowered to 20 GeV and the pseudo rapidity range increased to $|\eta|<2.4$ on particle level\footnote{We did not alter the requirements on muons at the detector level, as our calibration has only been validated for muons with $\pT$>25 GeV within $|\eta|$<2.1. The modelling uncertainties due to the extrapolation are assumed to be small.}. Moreover, at least one reconstructed jet with a transverse energy of at least 30 GeV within $|y^{jet}|<2.4$ at particle and detector level is required. Jets are also reconstructed using generator particles, by clustering final-state particles with decay length $c\tau$ > 10 mm, using the anti-k$_{\textrm{T}}$ algorithm with radius parameter R = 0.5. In total, 434179 candidate events in the electron decay channel and 473626 candidate events in the muon decay channel were selected, while 61447 events with more than one reconstructed jet were found. The number of selected events as well as the fiducial cross section definitions for all Drell-Yan measurements are summarized in Table~\ref{tab:EventsFiducial}. 

The selection of positively and negatively charged $W$ bosons is only applied in the muon decay channel, since no simulated $W^\pm\rightarrow e\nu$ samples at $\sqrt{s}=8\,$ TeV were available on the CERN Open Data Portal. The $W$ boson candidate events are selected at detector level by requiring exactly one reconstructed, tightly isolated muon with $\pT>$ 25 GeV and within $|\eta|<2.1$. A minimum \MET\,of 25 GeV is required and a minimum transverse mass requirement of $\mT = \sqrt{2 \cdot \pT^l \cdot \MET \cdot (1-\cos(\Delta \phi))}>40$ GeV is applied, where $\Delta \phi$ is the opening angle of the reconstructed muon and the missing transverse momentum vector in the transverse plane. This selection differs from the chosen approach in \cite{Chatrchyan:2014mua} and is closer to \cite{Aad:2019rou} in order to reduce multijet background contributions. Similar kinematic constraints are applied at particle level on the charged decay lepton, the neutrino and the derived quantities (Table \ref{tab:EventsFiducial}). In total, 3631170 $W^+$ and 2629480 $W^-$ candidate events have been selected, respectively.

The selection of top-quark pair events is performed only in the electron-muon final state, i.e. focusing on $t\bar t \rightarrow W^\pm b W^\mp \bar{b} \rightarrow (\mu^\pm \nu) b (e^\mp \nu) \bar{b} $ due to its small background contributions, using data that is triggered by the single muon trigger. Only events with exactly one loose isolated muon (within  $|\eta|<2.1$) and exactly one oppositely charged loose isolated electron (within $|\eta|<1.44$ or $1.57<|\eta|<2.5$) are selected. The minimum transverse energy/momentum requirements for both leptons is 25 GeV and the minimum \MET\, requirement is 40 GeV. Moreover, it is required that the candidate events contain at least two reconstructed jets with $\pT>40$ GeV within $|y|<2.4$.  At least one of the reconstructed jets on detector level in the event has to be b--tagged. The number of candidate events passing this selection is 1495. The requirements imposed at particle level are significantly loosened, e.g. no cuts on the neutrinos or jets are applied (see \ref{tab:EventsFiducial}).

The $WZ$ diboson production cross section is studied in the fully leptonic final state, i.e. requiring at least three charged, loosely isolated electrons or muons with $\pT>25$ GeV within $|\eta|<2.1$ at detector level, but within $|\eta|<2.5$ at particle level. We only use data that is triggered by the single-muon trigger as its performance could be cross-checked in the $W\rightarrow\mu\nu$ analysis. Hence, the $eee\nu$ final state is not considered further \footnote{No MC sample of the $W\rightarrow e\nu$ processes is available within the CMS Open Data initiative, which would allow the validation of the corresponding single electron trigger efficiency.}. The missing transverse energy at detector level is required to be larger than 20 GeV. The oppositely charged leptons of the same flavor whose invariant mass, $m_{ll}$, is closest to the $Z$ boson candidate mass are required to fulfill $66<m_{ll}<116$ GeV. The third lepton is identified as the $W$ boson decay lepton, and the resulting transverse mass is required to be above 40 GeV. Similar, but not exactly identical, requirements are applied at the particle level objects and summarized in Table \ref{tab:EventsFiducial} together with the number of selected $WZ$ candidate events in data. The efficiency correction factors $C$ for the seven different processes considered are also summarized in  Table \ref{tab:EventsFiducial}. 

\begin{table}[tb]
\footnotesize
\begin{center}
\begin{tabular}
{l | l | c | c}
\hline
\textbf{Process}									&	\# selected events	& Definition of fiducial phase-space		 								&	 $C$ factor			\\
\hline
$Z/\gamma^*\rightarrow e^+e^-$					&	434,179			& $(1e^+ 1e^-)$, $60<m_{ee}<120$ GeV, 										&	$0.525\pm0.015$	\\
											&					& $p^e_\textrm{T}>25$ GeV, $|\eta^e|<2.1$											&					\\
\hline
$Z/\gamma^*\rightarrow \mu^+\mu^-$				&	473,626			& $(1\mu^+1\mu^-)$, $60<m_{\mu\mu}<120$ GeV, 								&	$0.637\pm0.010$	\\
											&					& $p^\mu_\textrm{T}>25$ GeV, $|\eta^\mu|<2.1$										&					\\
\hline
$Z/\gamma^*\rightarrow \mu^+\mu^-$				&	61,447			& $(1\mu^+ 1\mu^-)$, $70<m_{\mu\mu}<110$ GeV, $p^\mu_\textrm{T}>20$ GeV,				&	$0.428\pm0.029$	\\
$+ \geq 1 jet$									&					& $|\eta^\mu|<2.4$, $p^{jet}_\textrm{T}>30$ GeV, $|y^{jet}|<2.4$ 							&					\\
											&					& $\Delta R(j,l)>0.5$														&					\\
\hline
$W^+\rightarrow \mu^+ \nu$ 						&	3,631,170			& $(1\mu^+)$, $p^\mu_\textrm{T}>25$ GeV, $|\eta^\mu|<2.4$,							&	$0.593\pm0.017$	\\
											&					& $p^\nu_\textrm{T}>25$ GeV, $m_\textrm{T}>40$ GeV										&					\\
\hline
$W^-\rightarrow \mu^- \nu$ 						&	2,629,480			& $(1\mu^-)$, $p^\mu_\textrm{T}>25$ GeV, $|\eta^\mu|<2.4$,							&	$0.611\pm0.018$	\\
											&					& $p^\nu_\textrm{T}>25$ GeV, $m_\textrm{T}>40$ GeV										&					\\
\hline
$t\bar{t}\rightarrow \mu^\mp e^\pm \nu \bar \nu b \bar b\,$	&	1495				& $1\mu^\pm, 1e^\mp$, $p^e_\textrm{T}, p^\mu_\textrm{T}>20$ GeV , $|\eta^e|,|\eta^\mu|<2.4$,						&	$0.177\pm0.012$				\\ 
\hline
$W^\pm $Z$ \rightarrow l^\pm \nu l^+ l^-\,$				&	79				& $(e^\pm e^\mp \mu^\pm), (\mu^\pm \mu^\mp e^\pm), (\mu^\pm \mu^\mp \mu^\pm)$,  $p^l_\textrm{T}>25$ GeV,  &$0.363\pm0.011$	\\
$(l=e.\mu)$									&					& $|\eta^l|<2.5$, $80<m_{ll}<100$ GeV, $m_\textrm{T}>40$ GeV								&					\\
\hline
\end{tabular}
\caption{Overview of selected candidate events, the definition of the corresponding fiducial phase-space regions as well as detector correction ($C$) factors for seven chosen validation processes.\label{tab:EventsFiducial}}
\end{center}
\end{table}

\subsection{\label{sec:backgrounds}Background Estimations}

The contribution of background processes other than multijet processes is estimated using fully simulated MC samples detailed in Table~\ref{tab:MCSamples}. Each of the signal selections is applied to those samples and the corresponding yields are evaluated and weighted by the corresponding cross section of the processes and data luminosity. Following previous analyses, we assume a conservative uncertainty of 5\% in the cross sections of all relevant background processes. The contributions of those background processes that are known to have only a small impact in the signal region, i.e. below the uncertainty of  the largest background contribution, are neglected.

Multijet backgrounds, as well as backgrounds involving non-prompt leptons or jets that are wrongly identified as leptons, are estimated in data. A so called $ABCD$ method is used where two orthogonal properties of events, separating signal from multijet background processes, are used to define four regions in phase space, of which one region ($A$) is the signal region. The events in regions B and C pass one signal selection criterion, but fail the second, while events in region D fail both signal selection requirements. Signal contributions, as well as contributions from background processes that have been determined via full MC simulations in the regions $B$, $C$ and $D$, are subtracted. Assuming no correlation between the two selection properties, the multi-jet background yield in region A can then be estimated by $N_A = N_B \cdot N_C / N_D$.

For Drell-Yan processes, the events are categorized as oppositely-charged and same-charged lepton pair events as well as in fully-isolated and semi- or non-isolated lepton pairs. When applying the $ABCD$ method described above, a multijet background contribution of 0.2\% is found.  A systematic 50\% systematic uncertainty is applied to this contribution. This is validated by varying the degree of non-isolation of lepton pairs and repeating the multijet background estimation. 

The definition of the $ABCD$ regions in $W^\pm$ boson processes are also isolated and non-isolated leptons, as well as events with ($\ET<$25 GeV, $\mT<$40 GeV) and ($\ET>$25 GeV, $\mT>$40 GeV), where a muon trigger without an isolation requirement has been used. This choice leads to a multijet background estimate of 200000 events. The systematic uncertainty in this value is estimated again by varying the degree of the lepton non-isolation as well as the requirements on the $\ET$ and $\mT$. An uncertainty of 30\% covers all observed variations in the background yield. The same regions are used for the $WZ$ diboson signal selection to estimate the multijet background as well as the Z+jets background contribution, where one jet is miss-identified as a lepton, yielding a background contribution of $<1\%$.

The multijet contribution in the $t\bar{t}$ study is estimated by studying events where both leptons fail the isolation requirement and/or fail the requirement on $\MET$, yielding to a relative contribution of below 1\%. A careful analysis of the multijet background can certainly reduce the corresponding systematic uncertainties for all estimations, however, the approach we chose is fully justified in the context of this study with its limited precision focus.

An overview of the expected background contributions is given in Table \ref{tab:Background}.

\begin{table}[tb]
\footnotesize
\begin{center}
\begin{tabular}
{l | c| c | c | c| c}
\hline
\textbf{Process}								& $Z\rightarrow\tau\tau$ & $Z\rightarrow\mu\mu$	&	\,\,\,\,$t\bar t\rightarrow 2l2\nu2b \,\,\,\,\,\,\,\,\,\,$		&	DiBoson		\,\,\,\,\,\,\,\,\,\,\,\,\,\,& \,\,\,\,\,\,\,Multijet	\,\,\,\,\,\,\,\\
											&				& &	$t\bar t\rightarrow 1l1\nu2b2q $						&		\\
\hline
$Z/\gamma^*\rightarrow e^+e^-$					&	<0.2\% & -	&	0.2\%						&	<0.1\%		& 0.2\%	\\
\hline
$Z/\gamma^*\rightarrow \mu^+\mu^-$				&	0.1\% & -&	0.1\%						&	<0.1\%		& <0.2\%	\\
\hline
$Z/\gamma^*\rightarrow \mu^+\mu^- +\geq 1 $ jet		&	0.1\% & -	&	0.6\%						&	0.7\%		& <0.2\%	\\
\hline
$W^+\rightarrow \mu^+ \nu$ 						&	- & 5.3\%	&	0.3\%						&	<0.1\%		& 2.8\%	\\
\hline
$W^-\rightarrow \mu^- \nu$ 						&	- & 5.8\%	&	0.4\%						&	<0.1\%		& 3.8\%	\\
\hline
$t\bar{t}\rightarrow \mu^\mp e^\pm \nu \bar \nu b \bar b\,$	&	0.8\% & -		&	4.0\%						&	<0.1\%		& <0.1\%	\\ 
\hline
$W^\pm $Z$ \rightarrow l^\pm \nu l^+ l^-\,(l=e.\mu)$		&	-	&-		&	-							&	8.1\%		& 1\%	\\
\hline
\end{tabular}
\caption{Overview of the relative contribution of background processes to the signal region\label{tab:Background}}
\end{center}
\end{table}

\subsection{\label{sec:sys}Systematic Uncertainties}

The systematic uncertainties in the detector correction factors $C$ (see Section \ref{Sec:Calibration}) have been evaluated within our analysis framework by varying each correction independently within its uncertainties. The difference of the resulting correction factor after a particular variation $i$, $C^\prime_i$, to the nominal $C$ factor, is then taken as systematic $\Delta C = C^\prime_i - C$. When applicable, these differences are symmetrized for up- and down-variations. The systematic uncertainties due to pile-up are estimated by comparing the selection with and without the $\rho$-parameter reweighting. All relevant sources of the uncertainties are treated as independent from each other and hence the total systematic uncertainty on $\Delta C_{tot}$ is given by the Gaussian sum of the individual uncertainties $\Delta C_i$. The systematic uncertainties in $C$ range between $3\%$ and $10\%$ and dominate over the statistical uncertainties due to the size of the MC samples. An overview of the uncertainty breakdown on the $C$ factors for all validation samples is given in Table~\ref{tab:Uncertainties}.

\begin{table}[tb]
\footnotesize
\begin{center}
\begin{tabular}{l | c|c| c|c| c|c|c|c| c}
\hline
\textbf{Process}									&Elec.		& Elec.			& Muon		& Muon			& JES/		& \,\,\,\MET\,\,\,	& b-tag-		& pile-	& Total\\
											&Eff. 		& Scale/			& Eff. 		& Scale/			& JER		& 			& ging 		& up		& 		\\
											&			& Res.			& 			& Res.			& 			& 			& 	 		&  		&		\\
\hline
$Z/\gamma^*\rightarrow e^+e^-$					&	2.9\%	& 0.2\%			&	-		&	-			&	-		&	-		&	-		&	0.1\%	&	2.9\%	\\
\hline
$Z/\gamma^*\rightarrow \mu^+\mu^-$				&	-		&	-			&	1.5\%	&	0.3\%		&	-		&	-		&	-		&	0.1\%	&	1.6\%	\\
\hline
$Z/\gamma^*\rightarrow \mu^+\mu^-+\geq 1 $ jet		&	-		&	-			&	1.5\%	&	0.4\%		&	6.5\%	&	-		&	-		&	0.3\%	&	6.7\%	\\
\hline
$W^+\rightarrow \mu^+ \nu$ 						&	-		&	-			&	0.8\%	&	0.2\%		&	-		&	2.2\%	&	-		&	1.8\%	&	2.9\%	\\
\hline
$W^-\rightarrow \mu^- \nu$ 						&	-		&	-			&	0.8\%	&	0.2\%		&	-		&	2.2\%	&	-		&	1.8\%	&	2.9\%	\\
\hline
$t\bar{t}\rightarrow \mu^\mp e^\pm \nu \bar \nu b \bar b\,$	&	1.5\%	&	0.2\%		&	1.0\%	&	0.2\%		&	5.2\%	&	1.4\%	&	3\%		&	1.9\%	&	6.7\%	\\ 
\hline
$W^\pm $Z$ \rightarrow l^\pm \nu l^+ l^-\,(l=e.\mu)$		&	1.3\%	&	0.2\%		&	1.5\%	&	0.3\%		&	-		&	1.3\%	&	-		&	1.8\%	&	3.0\%	\\
\hline
\end{tabular}
\caption{Relative uncertainties on the detector correction factor $C$ for all studied validation processes due to different systematic uncertainties of detector effects. The uncertainties on the electron efficiencies (Elec. Eff) as well as on the muon efficiencies (Muon. Eff.) summarize reconstruction, identification, isolation and trigger efficiencies. Scale and resolution effects (Scale/Res.) for electrons and muons, as well as jet energy scale and resolution uncertainties (JES/JER) are separated. \label{tab:Uncertainties}}
\end{center}
\end{table}

Numerous control distributions between data and MC for all seven processes have been validated. Good agreement between data and MC has been observed in all of them. The normalized invariant mass and lepton rapidity distributions for the Drell--Yan processes have been already discussed in Section \ref{Sec:Calibration}. Two selected jet distributions of the $Z$+jets study are shown as an example in Figure~\ref{fig:ZJetsPlots}, where good agreement of the MC prediction and the data can be observed. The measurement of the $W$ boson production cross section is able to test the description of \MET, hence Figure~\ref{fig:WPlots} displays the comparison of data and MC for \MET and \mT. Similarly, Figure \ref{fig:TTBar} shows the comparison of the leading jet \pT\, as well as \MET\,for the $t \bar t$ selection, with a similar conclusion. The study of $WZ$ diboson production is statistically limited. Control distributions of the invariant and transverse mass observables are shown in Figure~\ref{fig:WZ}, where a good agreement within the limited statistical precision is observed.

\begin{figure}[tb]
\begin{center}
\includegraphics[width=7.3cm]{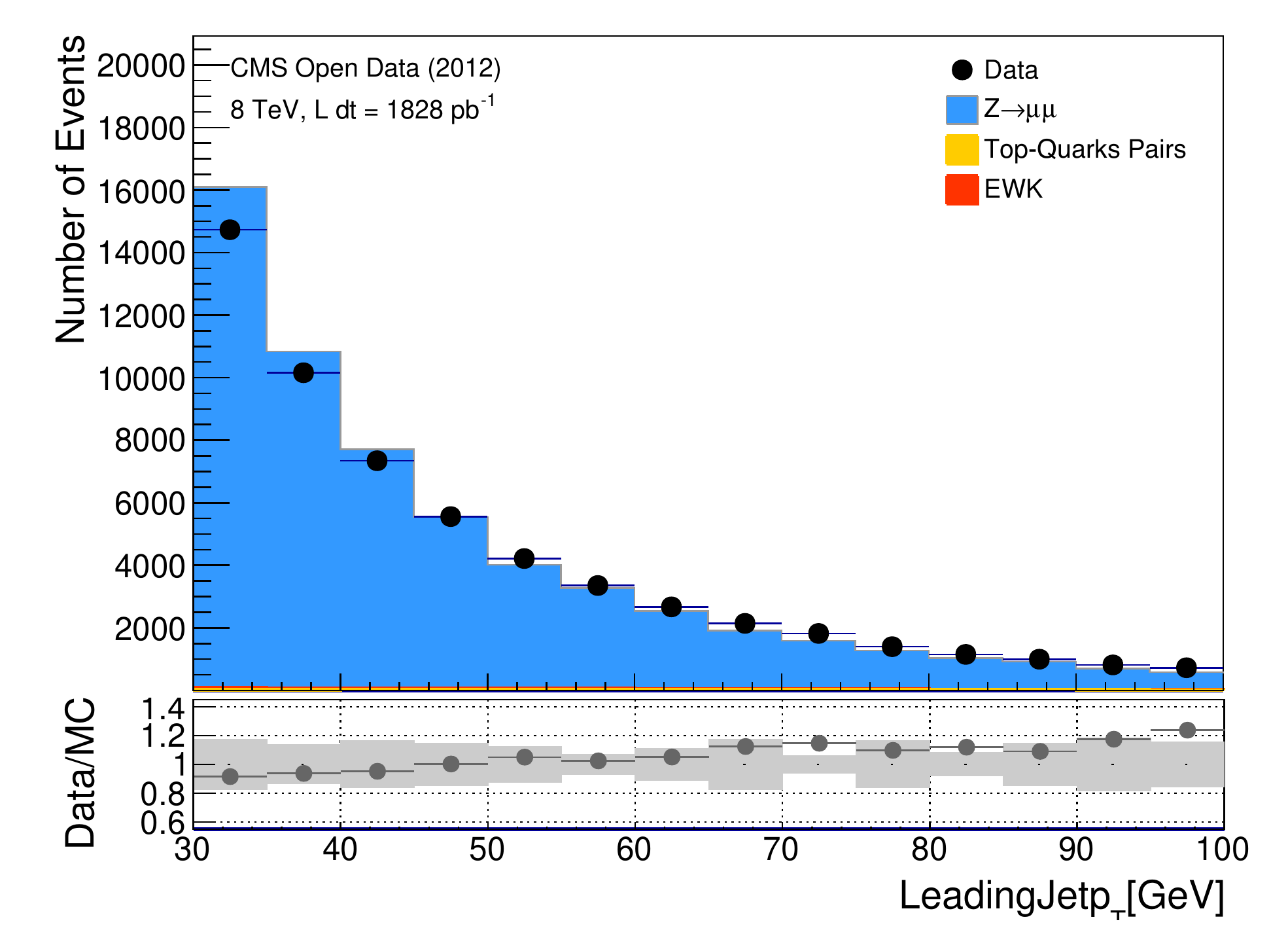} 
\hspace{0.1cm}
\includegraphics[width=7.3cm]{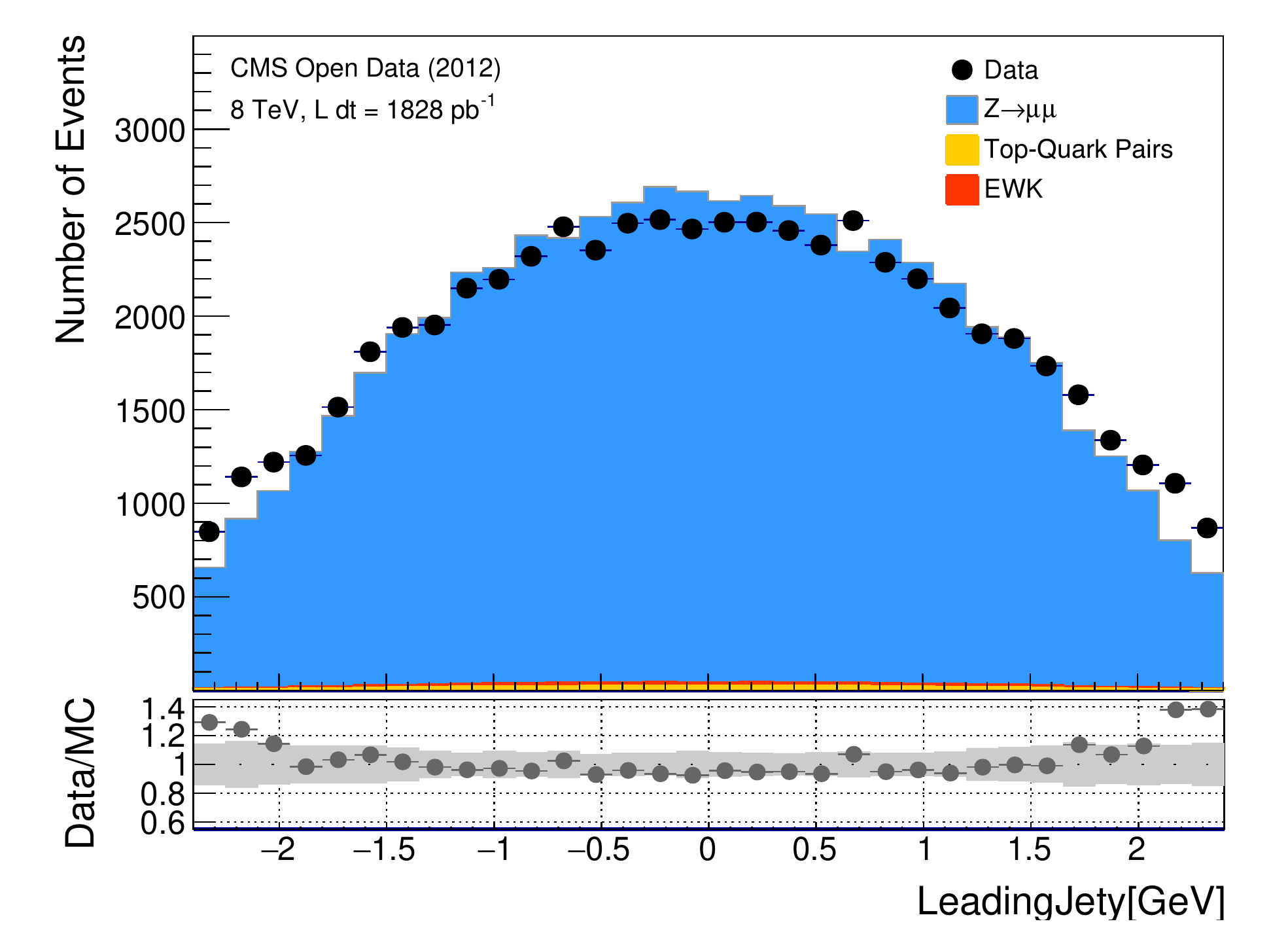}
\caption{\label{fig:ZJetsPlots} Normalized jet $\pT$ (left) and jet-rapidity distribution (right) for data and MC in the Z+jets study, after all corrections have been applied. The gray band in the ratio indicates the systematic uncertainties on the predictions without luminosity uncertainties.}
\end{center}
\end{figure}

\begin{figure}[tb]
\begin{center}
\includegraphics[width=7.3cm]{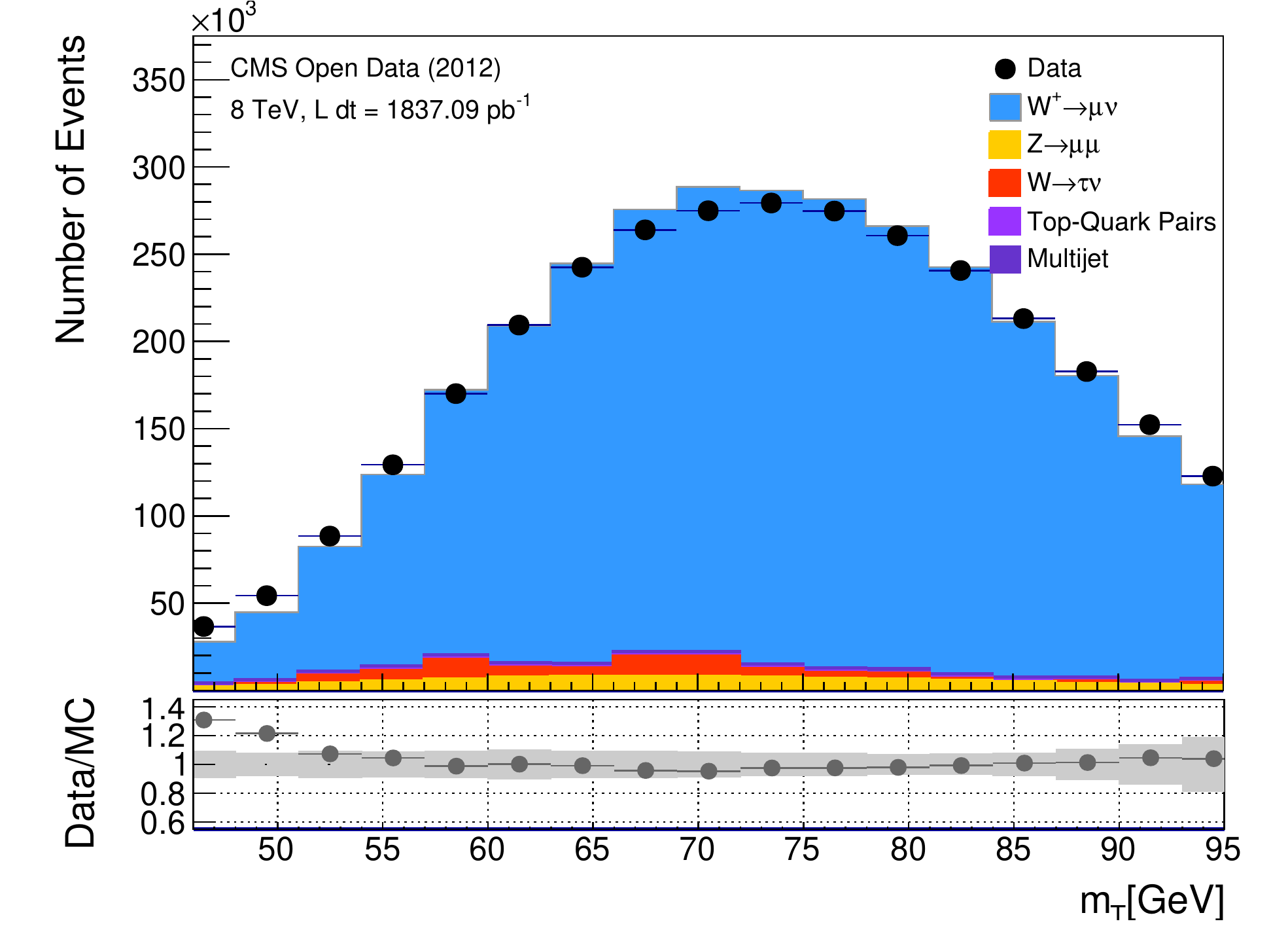} 
\hspace{0.1cm}
\includegraphics[width=7.3cm]{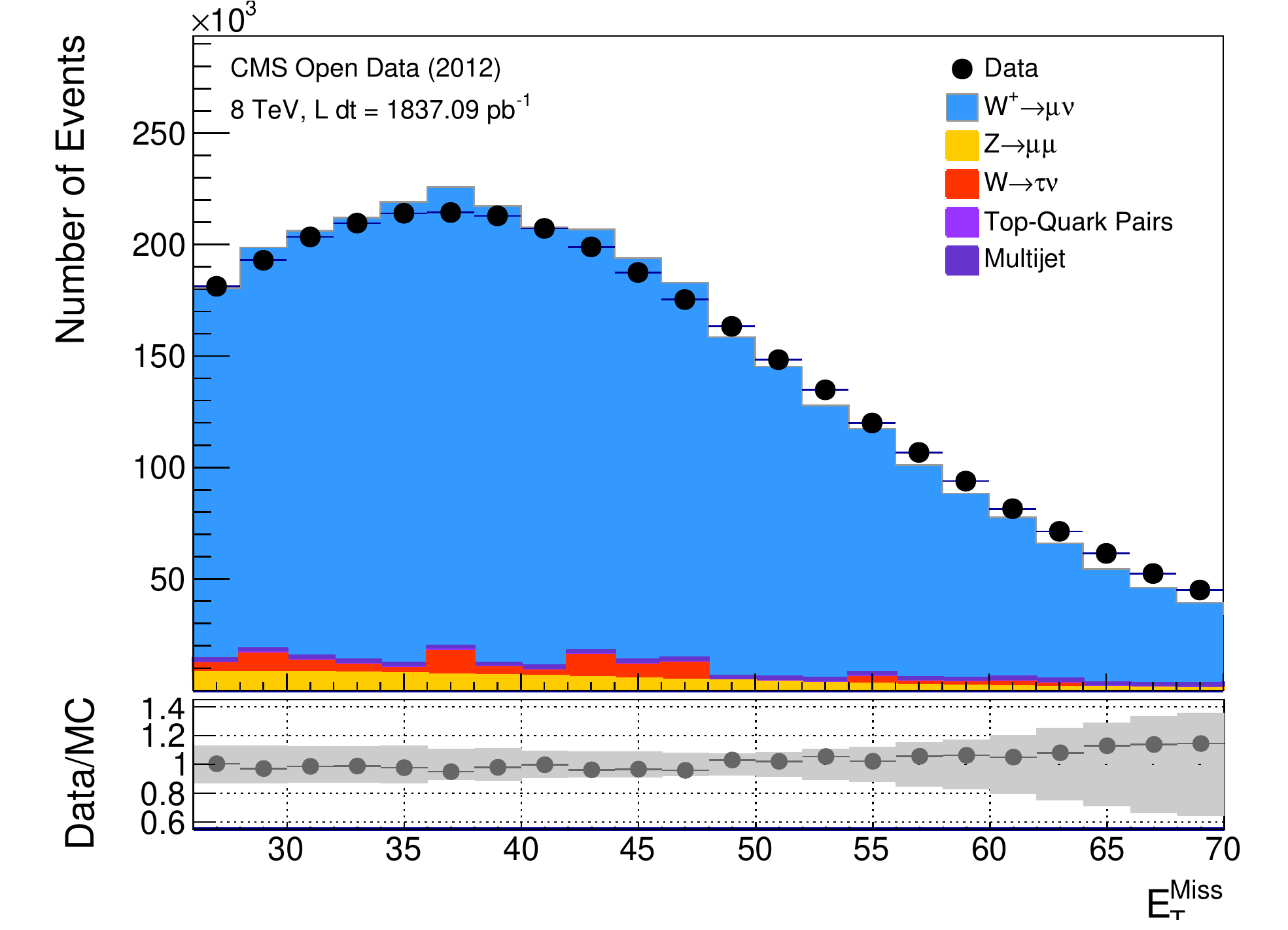}
\\
\includegraphics[width=7.3cm]{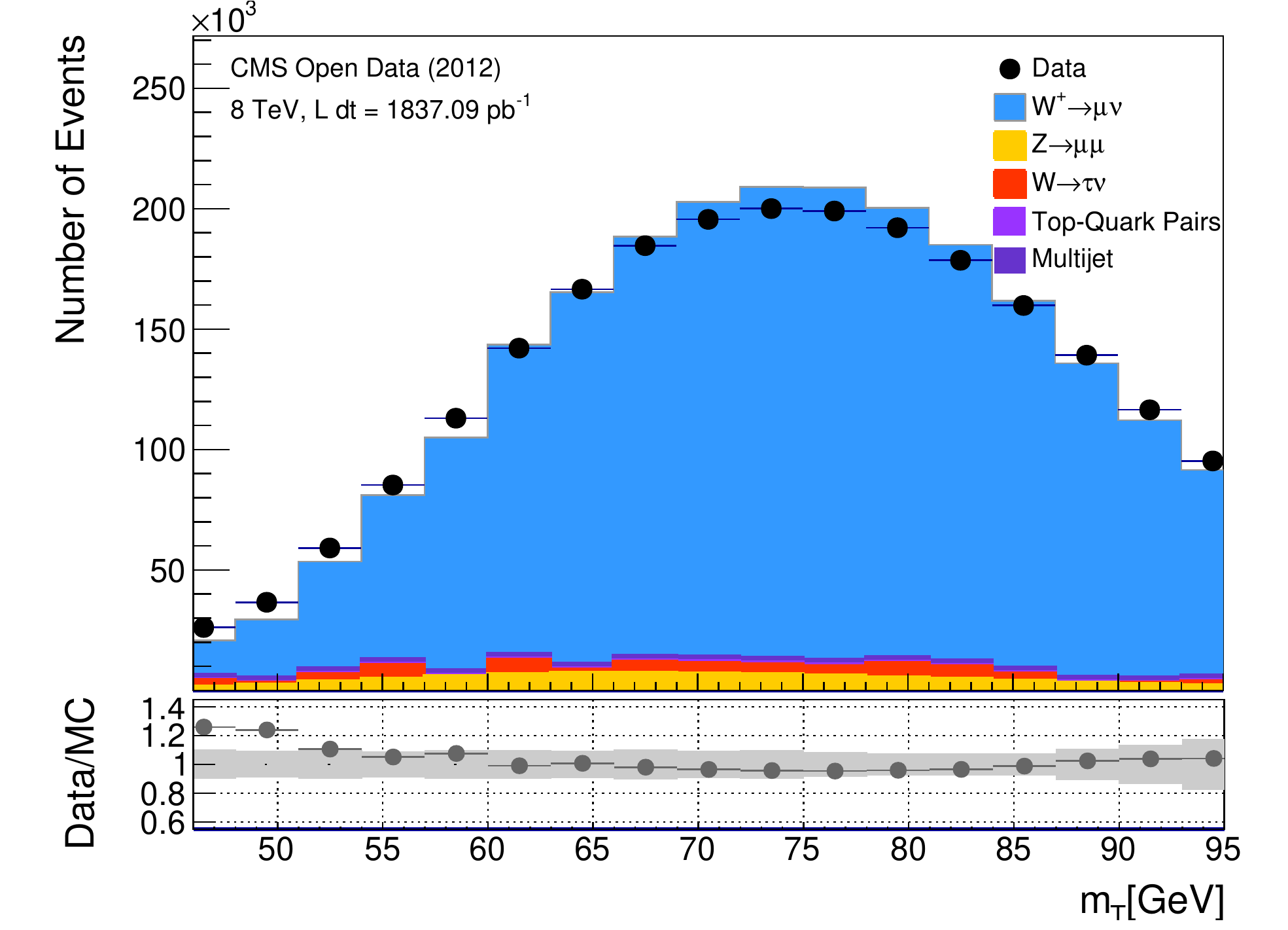} 
\hspace{0.1cm}
\includegraphics[width=7.3cm]{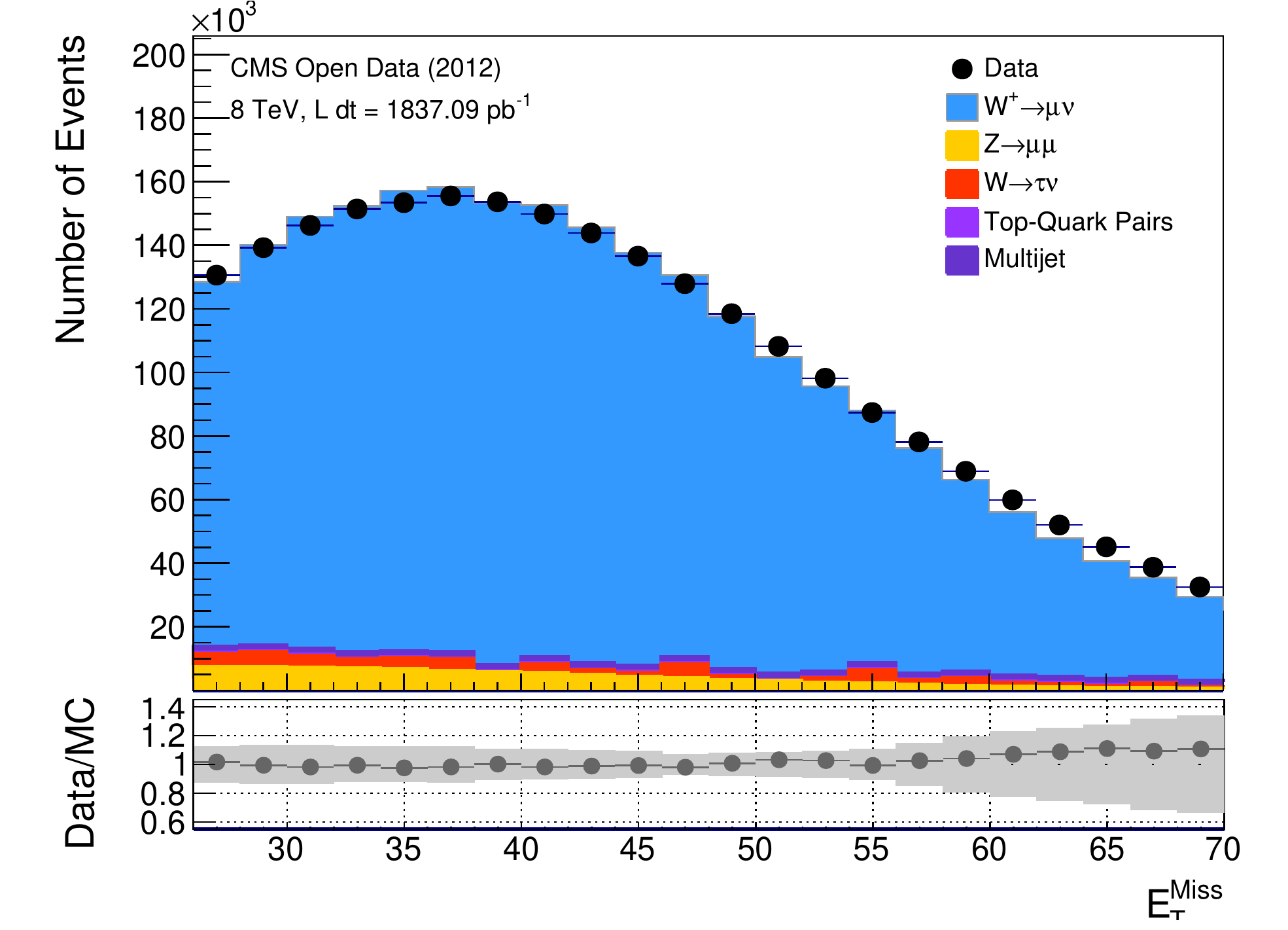}
\caption{\label{fig:WPlots} Normalized \mT~(left) and $\MET$ distribution (right) for data and MC in $W^+$ events (upper row) and $W^-$ events (lower row), after all corrections have been applied. The gray band in the ratio indicates the systematic uncertainties on the predictions without luminosity uncertainties.}
\end{center}
\end{figure}

\begin{figure}[tb]
\begin{center}
\includegraphics[width=7.3cm]{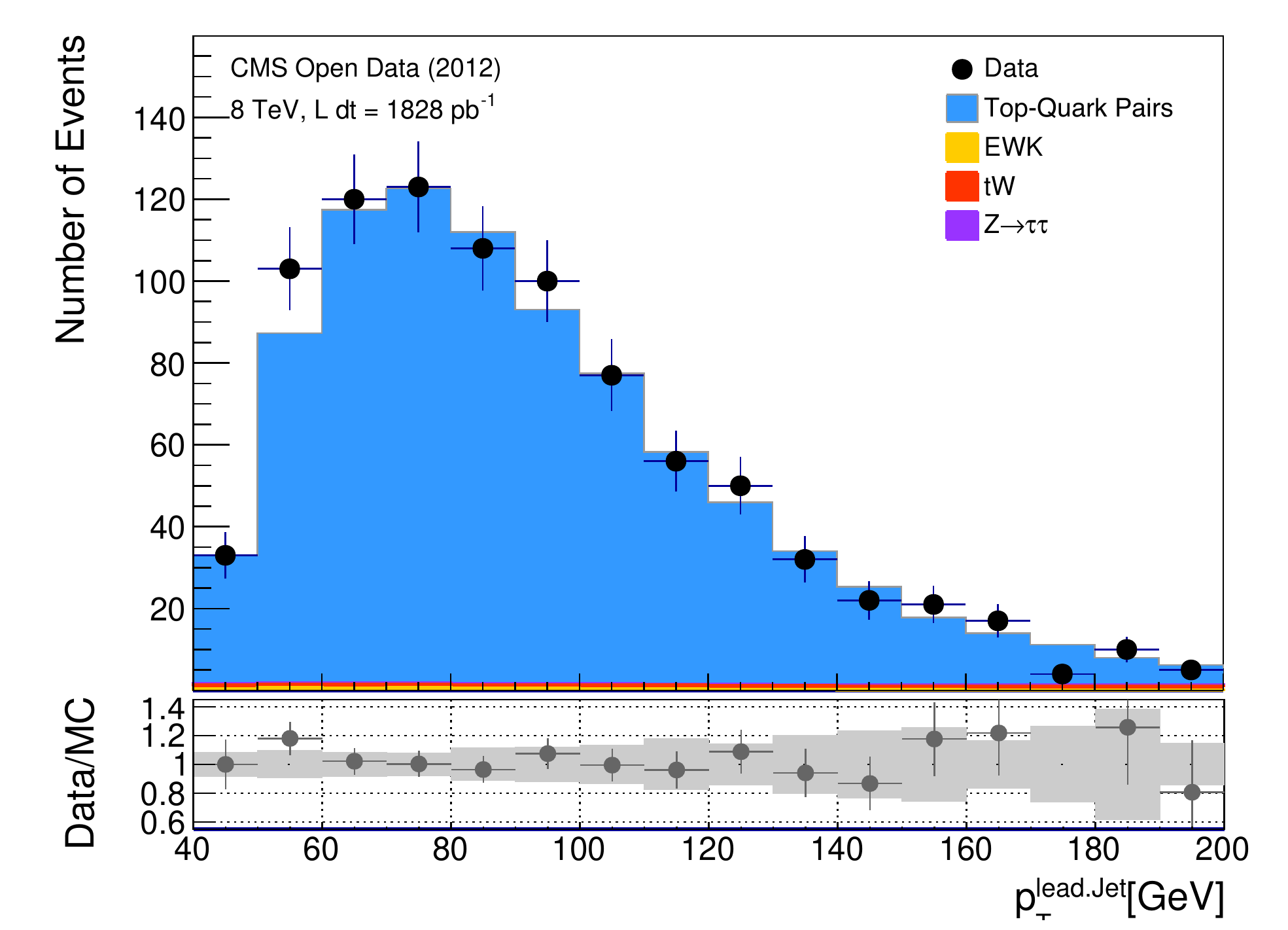} 
\hspace{0.1cm}
\includegraphics[width=7.3cm]{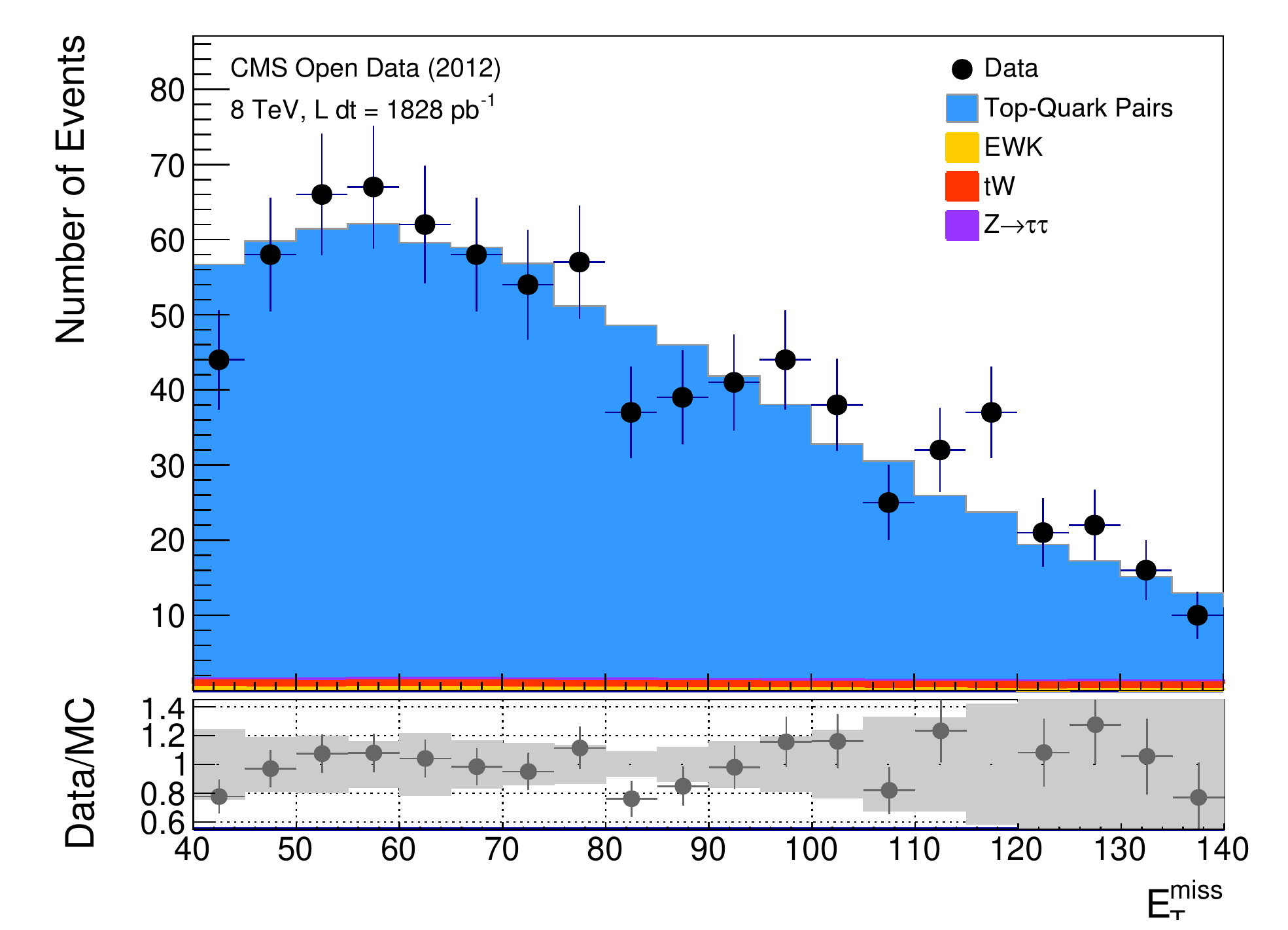}
\caption{\label{fig:TTBar} Normalized leading jet $\pT$ (left) and $\MET$ distribution (right) for data and MC in the $t\bar t$ study, after all corrections have been applied. The gray band in the ratio indicates the systematic uncertainties on the predictions without luminosity uncertainties.}
\end{center}
\end{figure}

\begin{figure}[tb]
\begin{center}
\includegraphics[width=7.3cm]{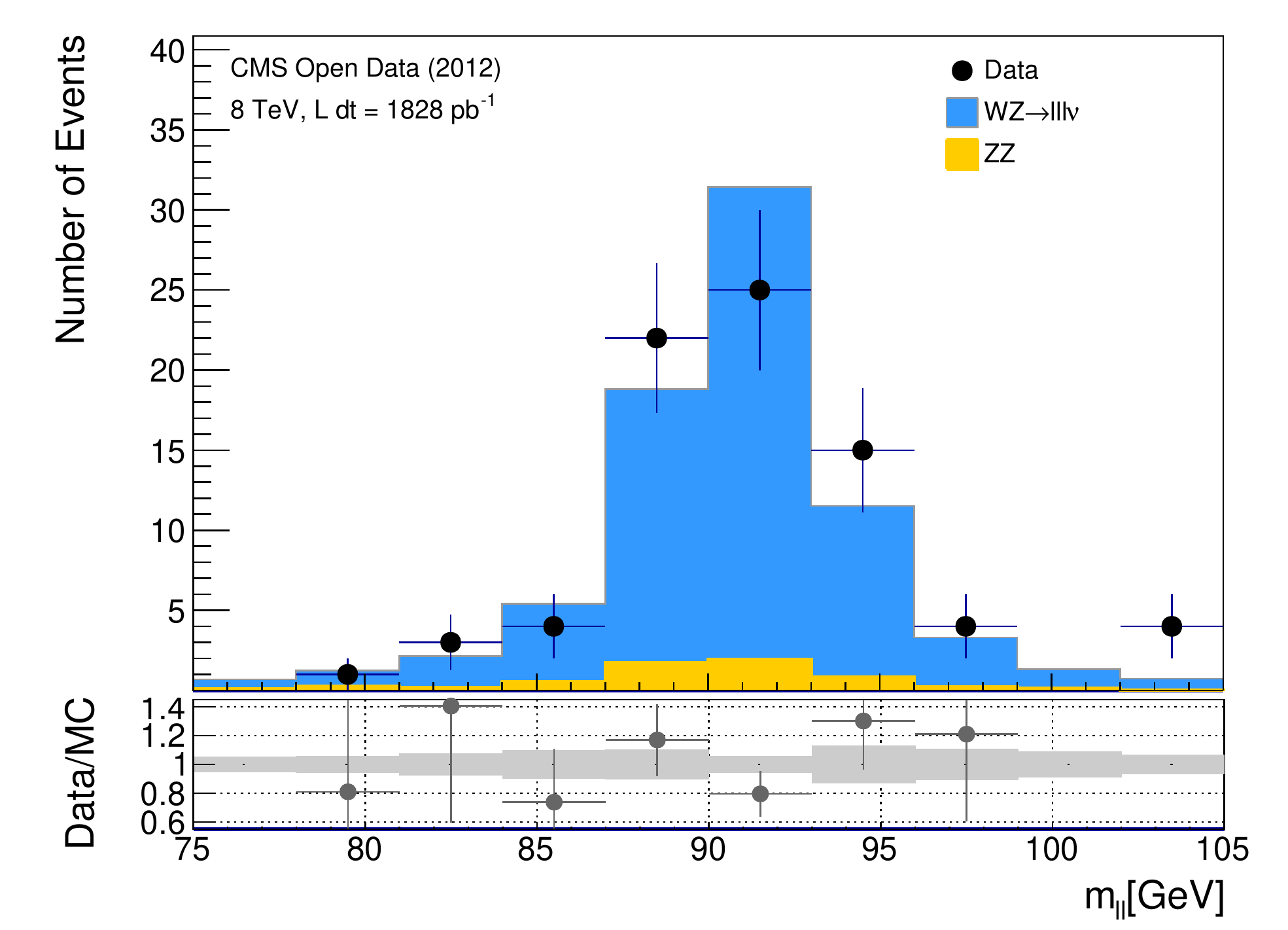} 
\hspace{0.1cm}
\includegraphics[width=7.3cm]{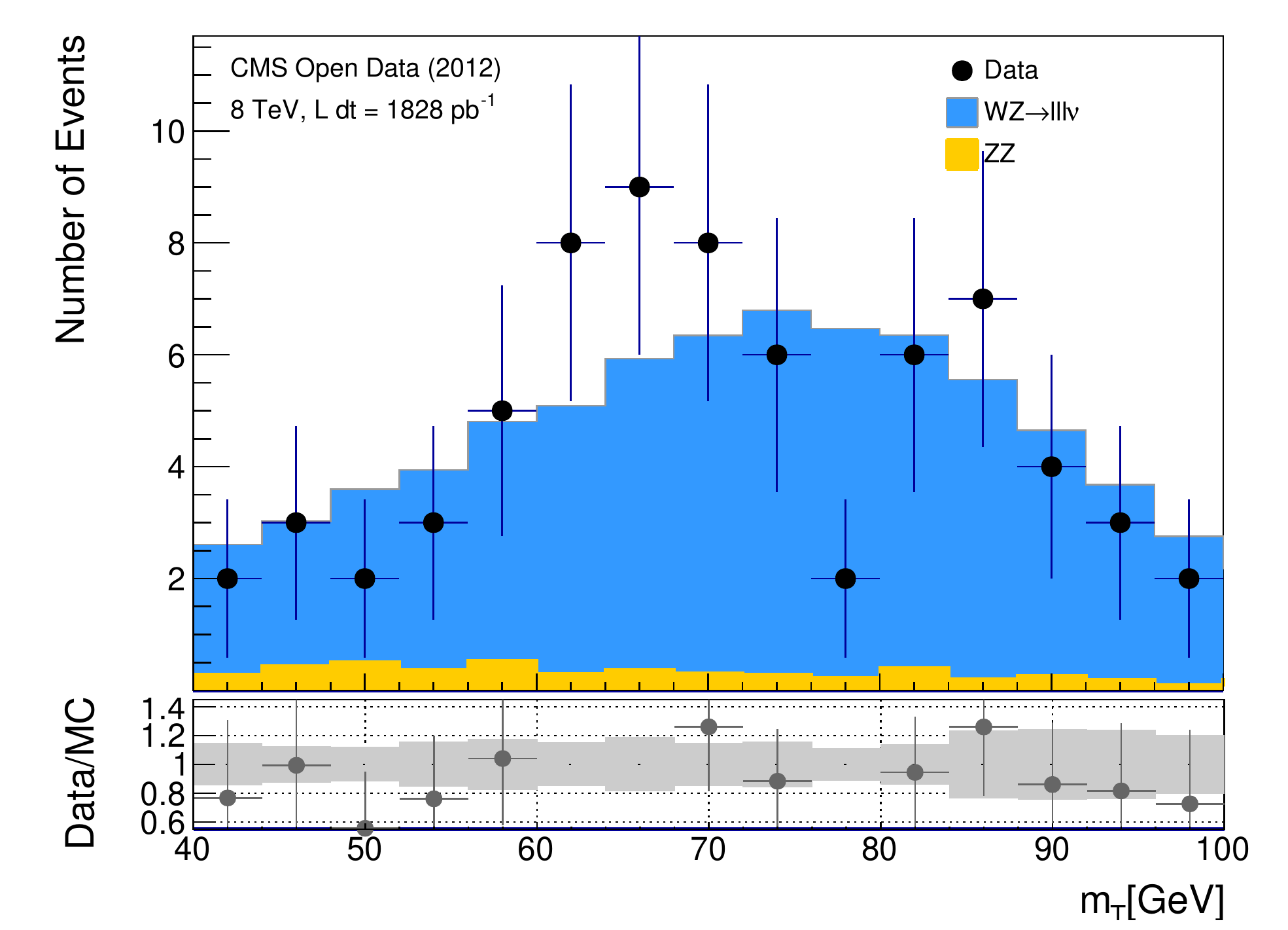}
\caption{\label{fig:WZ} Normalized invariant mass $m_{ll}$ (left) and transverse mass $\mT$ distribution (right) for data and MC in the $WZ$ study, after all corrections have been applied. The gray band in the ratio indicates the systematic uncertainties on the predictions without luminosity uncertainties.}
\end{center}
\end{figure}

\subsection{Results and Comparisons}

The fiducial production cross sections for the seven validation processes are determined via Eq. \ref{EQN:CrossSectionExp}, using the detector correction factors (Table \ref{tab:EventsFiducial}) as well as the estimated background contributions (Table \ref{tab:Background}). The resulting cross sections are summarized in Table \ref{tab:SMresults} and Figure \ref{fig:FigureSMSummary1}, together with their statistical and systematic uncertainties, where the latter are separated into detector-related and luminosity uncertainties. The results are also compared to previously published results from the CMS and ATLAS Collaborations, depending on which fiducial phase-space regions are closer to our own choices \cite{Chatrchyan:2014mua, Khachatryan:2016crw, Aad:2019rou, Khachatryan:2016mqs, Khachatryan:2016poo}. The $t\bar t$ and $WZ$ cross sections are compared to the published inclusive production cross sections \footnote{No model uncertainties were considered when extrapolating from our fiducial cross section to the inclusive cross section.}. Apart from the diboson $WZ$ process, all cross section measurements are dominated by systematic uncertainties due to the detector effects. The uncertainties due the background processes are small for all channels, except the $W^\pm$ boson production, where the uncertainty in the multijet background is of a similar size as the uncertainties due to the missing transverse energy requirements. 
It should be noted that the uncertainties for the $Z/\gamma^*\rightarrow l^+l^-$ processes in our analysis appear to be smaller than the official measurements by CMS \cite{Chatrchyan:2014mua}, however, this is due to the smaller data-set used by CMS as well as the limited number of significant digits in the published result. The smaller systematic uncertainties in the diboson $WZ$ cross section measurement is due to the smaller lepton reconstruction uncertainties assumed in our analysis and motivated in Section \ref{Sec:Calibration}.  We also compare the inclusive cross section to the available theoretical predictions, which have been previously published. Figure \ref{fig:FigureSMSummary2} shows the ratio of the theory predictions to our measured fiducial cross sections and to the previously published results by either ATLAS or CMS. All our measurements are in good agreement with the previously published results as well as the SM predictions. The measurement systematic uncertainties are between 1.6 and 6.7\%. 

\begin{table}[tb]
\footnotesize
\begin{center}
\begin{tabular}{l | c|c| c}
\hline
\textbf{Process}									& Measurement			&  Prediction					& \,\,\,\,\,\,\,\,\,\,\,\,\,\,Previous Result 		 	\\
											& cross section [pb]			&  cross section [pb]				& \,\,\,\,\,\,\,\,\,\,\,\,\,\, cross section [pb]	 	\\
											& $(stat.\pm sys. \pm lumi.)$	&  		  					& \,\,\,\,\,\,\,\,\,\,\,\,\,\,$(stat.\pm sys. \pm lumi.)$	\\
\hline
$Z/\gamma^*\rightarrow e^+e^-$					& $\sigma^{fid}=461\pm17$	&  $\sigma^{fid.}=450\pm20$ \cite{Chatrchyan:2014mua}		&  $\sigma^{fid}=450\pm20$ 	\cite{Chatrchyan:2014mua}\\
											& $(1\pm13\pm11)$			&  							&  $(10\pm10\pm10)$		\\
\hline
$Z/\gamma^*\rightarrow \mu^+\mu^-$				& $\sigma^{fid}=406\pm12$	&  $\sigma^{fid.}=400\pm10$ \cite{Chatrchyan:2014mua}		&  $\sigma^{fid}=410\pm20$ 	\cite{Chatrchyan:2014mua}\\
											& $(1\pm6\pm10)$			& 							&  $(10\pm10\pm10)$		\\
\hline
$Z/\gamma^*\rightarrow \mu^+\mu^-+\geq 1 $ jet		& $\sigma^{fid}=77.1\pm5.5$	&  $\sigma^{fid.}=76.3\pm5.0$	\cite{Khachatryan:2016crw}		&  $\sigma^{fid}=75.5\pm4.0$ 	\cite{Khachatryan:2016crw}\\
											& $(0.4\pm5.1\pm1.9)$		&  							&  $(0.1\pm 3.7\pm1.4)$		\\
\hline
$W^+\rightarrow \mu^+ \nu$						& $\sigma^{fid}=3052\pm124$	&  $\sigma^{fid.}=3015\pm100$	 \cite{Aad:2019rou}	&  $\sigma^{fid}=3110\pm66$  \cite{Aad:2019rou}	\\
											& $(1\pm98\pm76)$			& 							&  $(0.5\pm29\pm59)$		\\
\hline
$W^-\rightarrow \mu^- \nu$						& $\sigma^{fid}=2103\pm86$	&  $\sigma^{fid.}=2105\pm60$ \cite{Aad:2019rou}		&  $\sigma^{fid}=2137\pm47$ 	\cite{Aad:2019rou}\\
											& $(1\pm69\pm52)$			& 							&  $(0.4\pm22\pm41)$		\\
\hline
$t\bar{t}\rightarrow \mu^\mp e^\pm \nu \bar \nu b \bar b\,$	& $\sigma^{incl.}=4.54\pm0.35$	&  $\sigma^{incl.}=4.37\pm0.35$ \cite{Khachatryan:2016mqs}		&  $\sigma^{incl.}=4.23\pm0.14$ 	\cite{Khachatryan:2016mqs}\\
											& $(0.14\pm0.30\pm0.11)$	& 							&  $(0.02\pm0.10\pm0.10)$	\\
\hline
$W^\pm $Z$ \rightarrow l^\pm \nu l^+ l^-$				& $\sigma^{incl.}=28.1\pm3.3$	&  $\sigma^{incl.}=23.7\pm0.4$	\cite{Khachatryan:2016poo}	&  $\sigma^{incl.}=24.09\pm1.8$ 	\cite{Khachatryan:2016poo}\\
											& $(3.1\pm0.9\pm0.7)$		& 							&  $(0.87\pm1.6\pm0.6)$		\\
\hline
\end{tabular}
\caption{Overview of measured cross sections of seven validation processes as well as previously published results together with theory predictions.\label{tab:SMresults}}
\end{center}
\end{table}

\begin{figure}[t]
\centering
\begin{minipage}{7.3cm}
  \centering
  \includegraphics[width=1.0\linewidth]{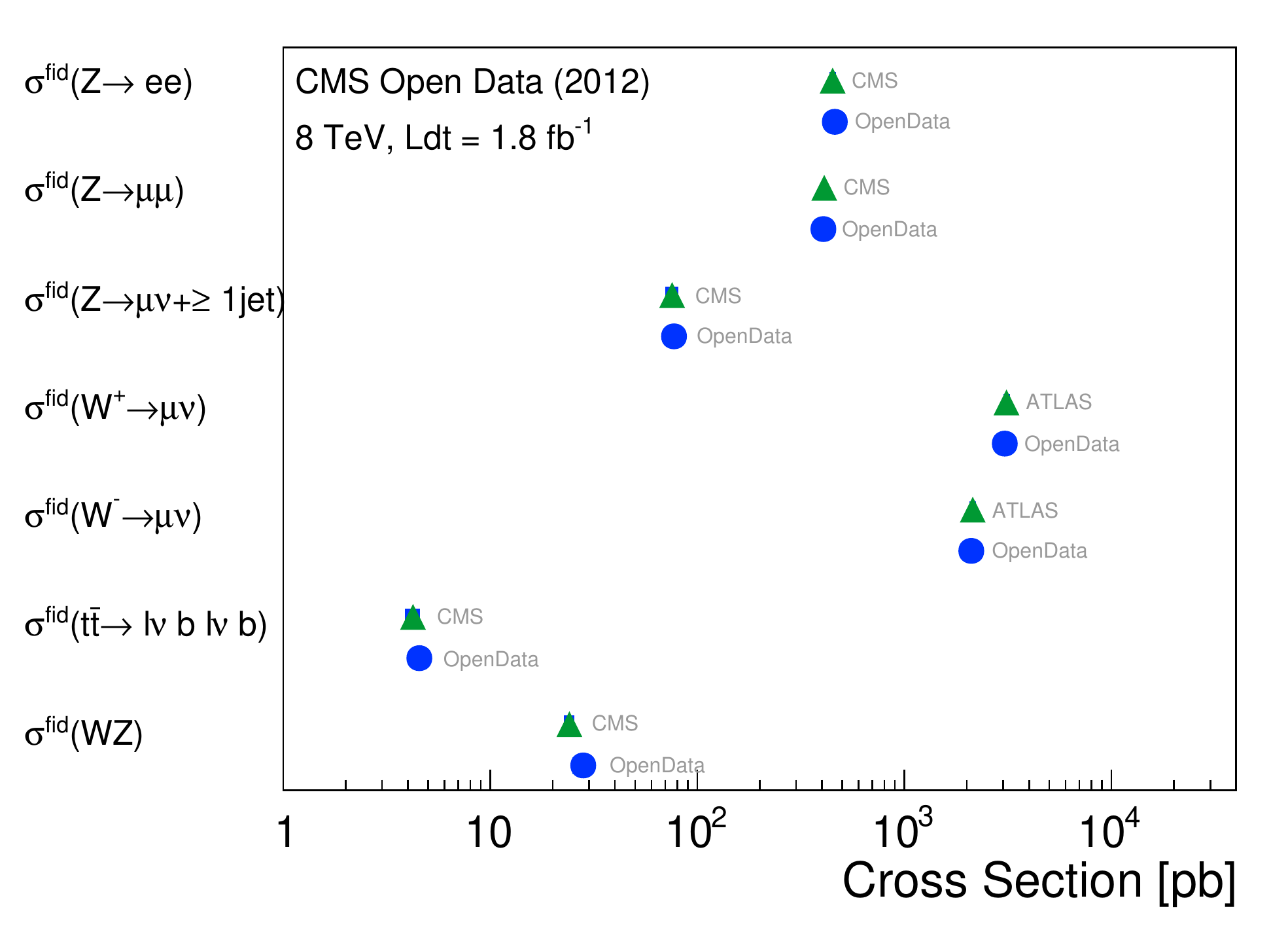}
	\caption{Overview of the measured SM production cross sections of seven validation processes. \label{fig:FigureSMSummary1}}
\end{minipage}%
\hspace{0.1cm}
\begin{minipage}{7.3cm}
\centering
\includegraphics[width=1.0\linewidth]{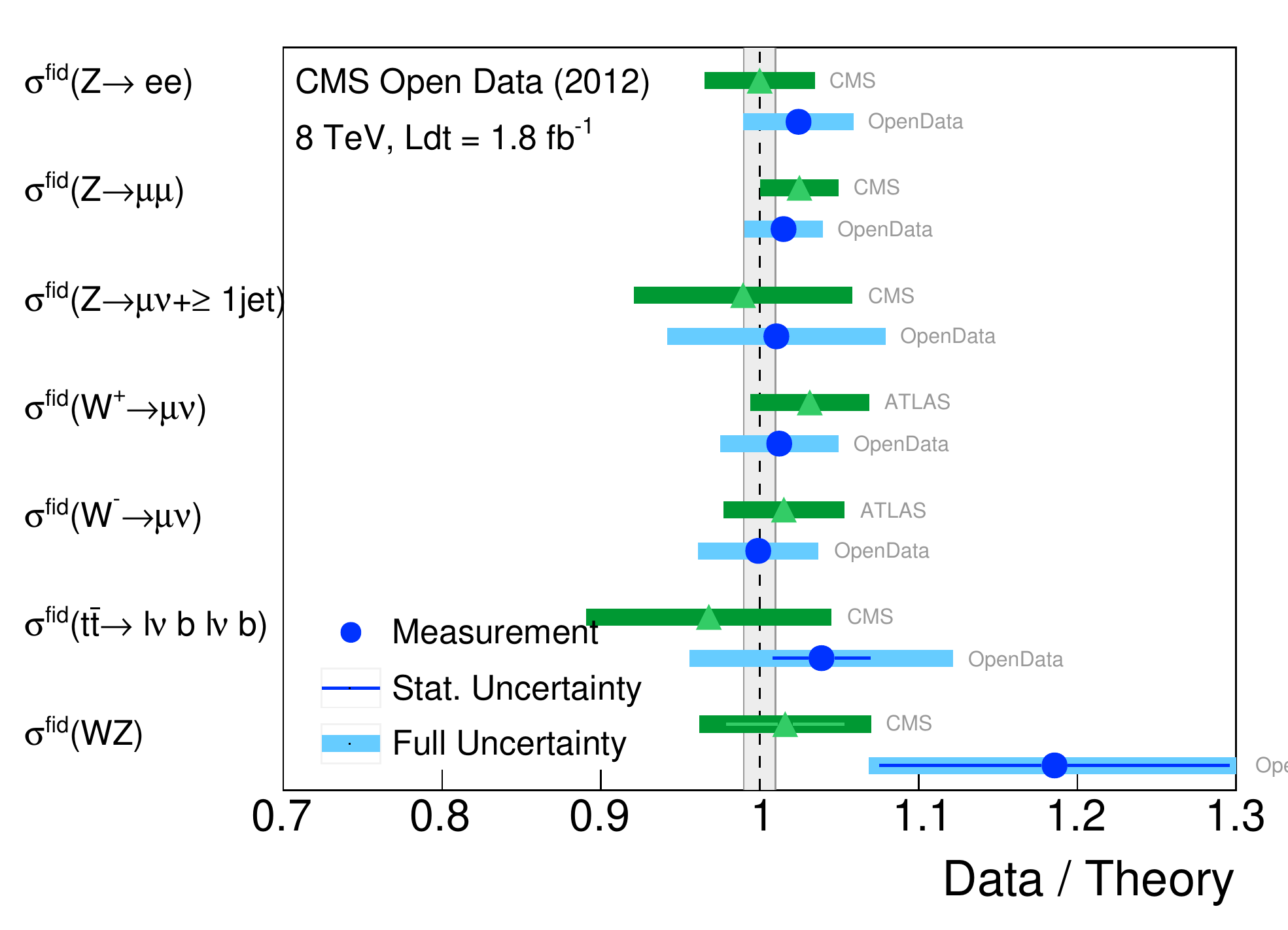}
\caption{Ratio of measured cross sections, derived in this work as well as by the LHC collaborations, to their corresponding theoretical predictions. \label{fig:FigureSMSummary2}}
\end{minipage}
\end{figure}


\section{\label{Sec:Conclusion}Opportunities and Challenges of the CMS Open Data Initiative}

The CMS Open Data initiative offers a unique opportunity to study and measure properties of the SM as long as a limited precision is sufficient. Measurements with higher precision currently seem not achievable, given the limited available information on the detector calibration as well as the systematic uncertainties of relevant observables. Clearly, these calibration efforts are one of the main areas of research within the Collaborations and the publications of the corresponding information in an easily accessible and understandable format for external physicists is highly challenging. One example is that experimental uncertainties in the energy scale of particle jets involve dozens of calibration parameters which have to be applied correctly. Another example are uncertainties in the lepton identification efficiencies, which are correlated in a complex manner, where the correlations matter for high precision measurements. Hence, precision measurements of (differential) cross sections as well as cross section ratios should, and can only, be performed by the LHC Collaborations. 

As a first possible improvement of the CMS Open Data initiative, we suggest that some simplified baseline calibrations, as well as uncertainties, should become available. This is highly desirable to make legacy data analyses possible to confront future theories with data. As a second possible improvement of the CMS Open Data initiative, we suggest that the CMS Collaboration publish dedicated baseline analyses, such as presented in Section \ref{Sec:BasicMeasurements}. This would allow external physicists to understand the technical details, e.g. how to apply the detector calibration or simply how to calculate the integrated luminosity for a given data sample.

However, even given the mentioned shortcomings, we see a significant physics potential in the CMS Open Data initiative as illustrated in this work. We have demonstrated that it is possible to repeat measurements of the $Z$ boson production cross section in the electron and muon decay channels, also in association with one jet in the final state, of the $W^\pm$ boson production cross section in the muon decay channel, of the top-quark pair production cross section in the fully leptonic decay mode, and of the $WZ$ diboson production cross section. Our validation measurements agree within less than 3\% to the official measurements by the CMS and ATLAS Collaborations. The differences are within the statistical and systematic uncertainties of the measurements. This lays the foundation to extend cross section measurements to extreme phase-space regions, which have not been probed so far.  


\acknowledgments

We would like to thank the CMS Collaboration for providing the full 2012 data set as well as for the documentation on the CMS detector performance. This work would have not been possible without the excellent performance of the LHC as well as the existing computing infrastructure and the support from CERN. We would also like to thank Guillelmo Gomez-Ceballos and Frank Fiedler for the helpful comments during the revision of this paper. M.S. would like to thank in addition the Fulbright commission as well as the Volkswagen Foundation for the support of this work. Moreover, he would like to thank his colleagues at MIT, Philip Harris in particular, for answering all questions regarding the treatment of the CMS Open Data for this project as well as the pleasant environment during the Fulbright research scholarship. 

\bibliographystyle{unsrt}
\bibliography{paper}

\begin{thebibliography}{10}

\bibitem{Aaboud:2016btc}
{Precision measurement and interpretation of inclusive $W^+$ , $W^-$ and
  $Z/\gamma ^*$ production cross sections with the ATLAS detector}.
\newblock {\em Eur. Phys. J.}, C77(6):367, 2017.

\bibitem{Chatrchyan:2014mua}
{Measurement of inclusive W and Z boson production cross sections in pp
  collisions at $\sqrt{s}$ = 8 TeV}.
\newblock {\em Phys. Rev. Lett.}, 112:191802, 2014.

\bibitem{Chatrchyan:2013faa}
{Measurement of the $t \bar{t}$ production cross section in the dilepton
  channel in pp collisions at $\sqrt{s}$ = 8 TeV}.
\newblock {\em JHEP}, 02:024, 2014.
\newblock [Erratum: JHEP02,102(2014)].

\bibitem{Aad:2016izn}
{Measurement of the angular coefficients in $Z$-boson events using electron and
  muon pairs from data taken at $\sqrt{s}=8$ TeV with the ATLAS detector}.
\newblock {\em JHEP}, 08:159, 2016.

\bibitem{Aad:2015auj}
{Measurement of the transverse momentum and $\phi ^*_{\eta }$ distributions of
  Drell-Yan lepton pairs in proton-proton collisions at $\sqrt{s}=8$  TeV with
  the ATLAS detector}.
\newblock {\em Eur. Phys. J.}, C76(5):291, 2016.

\bibitem{CMS:OpenData}
Software framework for cms open data analysis.
\newblock \url{http://opendata.cern.ch/docs/about-cms}.
\newblock (2017).

\bibitem{Cesarotti:2019nax}
Cari Cesarotti, Yotam Soreq, Matthew~J. Strassler, Jesse Thaler, and Wei Xue.
\newblock {Searching in CMS Open Data for Dimuon Resonances with Substantial
  Transverse Momentum}.
\newblock {\em Phys. Rev.}, D100(1):015021, 2019.

\bibitem{Tripathee:2017ybi}
Aashish Tripathee, Wei Xue, Andrew Larkoski, Simone Marzani, and Jesse Thaler.
\newblock {Jet Substructure Studies with CMS Open Data}.
\newblock {\em Phys. Rev.}, D96(7):074003, 2017.

\bibitem{Chatrchyan:2008aa}
{The CMS Experiment at the CERN LHC}.
\newblock {\em JINST}, 3:S08004, 2008.

\bibitem{CMS-PAS-PFT-09-001}
{Particle-Flow Event Reconstruction in CMS and Performance for Jets, Taus, and
  MET}.
\newblock Technical Report CMS-PAS-PFT-09-001, CERN, Geneva, Apr 2009.

\bibitem{Adam:815410}
Wolfgang Adam, R~Fruehwirth, Are Strandlie, and T~Todor.
\newblock {Reconstruction of Electrons with the Gaussian-Sum Filter in the CMS
  Tracker at the LHC}.
\newblock Technical Report CMS-NOTE-2005-001, CERN, Geneva, Jan 2005.

\bibitem{Chatrchyan:2013dga}
{Energy Calibration and Resolution of the CMS Electromagnetic Calorimeter in
  $pp$ Collisions at $\sqrt{s} = 7$ TeV}.
\newblock {\em JINST}, 8:P09009, 2013.
\newblock [JINST8,9009(2013)].

\bibitem{Khachatryan:2015hwa}
{Performance of Electron Reconstruction and Selection with the CMS Detector in
  Proton-Proton Collisions at 8 TeV}.
\newblock {\em JINST}, 10(06):P06005, 2015.

\bibitem{CMS-PAS-MUO-10-002}
{Performance of muon identification in pp collisions at 7 TeV}.
\newblock Technical Report CMS-PAS-MUO-10-002, CERN, Geneva, 2010.

\bibitem{CMS-PAS-PFT-10-001}
{Commissioning of the Particle-flow Event Reconstruction with the first LHC
  collisions recorded in the CMS detector}.
\newblock Technical Report CMS-PAS-PFT-10-001, 2010.

\bibitem{Chatrchyan:2011ds}
{Determination of Jet Energy Calibration and Transverse Momentum Resolution in
  CMS}.
\newblock {\em JINST}, 6:P11002, 2011.

\bibitem{Chatrchyan:2012jua}
{Identification of b-quark jets with the CMS experiment}.
\newblock {\em JINST}, 8:P04013, 2013.

\bibitem{cit:bacon}
Caterina~Vernieri Philip~Harris, Cristina Ana Mantilla~Suarez et~al.
\newblock Bacon analysis framework.
\newblock \url{https://github.com/ksung25/BaconProd/tree/Run1}.
\newblock (2015).

\bibitem{Hinzmann:2011zz}
Andreas Hinzmann.
\newblock {Tools for physics analysis in CMS}.
\newblock {\em J. Phys. Conf. Ser.}, 331:032042, 2011.

\bibitem{CMS:GoodRunList}
{CMS list of validated runs for primary datasets of 2012 data taking. CERN Open
  Data Portal}.
\newblock Technical report.

\bibitem{Brun:1997pa}
R.~Brun and F.~Rademakers.
\newblock {ROOT: An object oriented data analysis framework}.
\newblock {\em Nucl. Instrum. Meth.}, A389:81--86, 1997.

\bibitem{CMS:MuData1}
Single mu primary dataset in aod format from run of 2012
  (/singlemu/run2012c-22jan2013-v1/aod).
\newblock CERN Open Data Portal: \url{http://opendata.cern.ch}.
\newblock DOI:10.7483/OPENDATA.CMS.9A4E.7SIR.

\bibitem{CMS:MuData2}
Single mu primary dataset in aod format from run of 2012
  (/singlemu/run2012b-22jan2013-v1/aod).
\newblock CERN Open Data Portal: \url{http://opendata.cern.ch}.
\newblock DOI:10.7483/OPENDATA.CMS.IYVQ.1J0W.

\bibitem{CMS:EData1}
Single electron primary dataset in aod format from run of 2012
  (/singleelectron/run2012c-22jan2013-v1/aod).
\newblock CERN Open Data Portal: \url{http://opendata.cern.ch}.
\newblock DOI:10.7483/OPENDATA.CMS.BAKP.W6TP.

\bibitem{CMS:EData2}
Single electron primary dataset in aod format from run of 2012
  (/singleelectron/run2012b-22jan2013-v1/aod).
\newblock CERN Open Data Portal: \url{http://opendata.cern.ch}.
\newblock DOI:10.7483/OPENDATA.CMS.8XN1.J5N7.

\bibitem{CMS:2013gfa}
{CMS Luminosity Based on Pixel Cluster Counting - Summer 2013 Update}.
\newblock 2013.

\bibitem{Alioli:2008gx}
Simone Alioli, Paolo Nason, Carlo Oleari, and Emanuele Re.
\newblock {NLO vector-boson production matched with shower in POWHEG}.
\newblock {\em JHEP}, 07:060, 2008.

\bibitem{Alioli:2010xd}
Simone Alioli, Paolo Nason, Carlo Oleari, and Emanuele Re.
\newblock {A general framework for implementing NLO calculations in shower
  Monte Carlo programs: the POWHEG BOX}.
\newblock {\em JHEP}, 06:043, 2010.

\bibitem{Sjostrand:2006za}
Torbjorn Sjostrand, Stephen Mrenna, and Peter~Z. Skands.
\newblock {PYTHIA 6.4 Physics and Manual}.
\newblock {\em JHEP}, 05:026, 2006.

\bibitem{Alwall:2011uj}
Johan Alwall, Michel Herquet, Fabio Maltoni, Olivier Mattelaer, and Tim
  Stelzer.
\newblock {MadGraph 5 : Going Beyond}.
\newblock {\em JHEP}, 06:128, 2011.

\bibitem{Gao:2013xoa}
Jun Gao, Marco Guzzi, Joey Huston, Hung-Liang Lai, Zhao Li, Pavel Nadolsky, Jon
  Pumplin, Daniel Stump, and C.~P. Yuan.
\newblock {CT10 next-to-next-to-leading order global analysis of QCD}.
\newblock {\em Phys. Rev.}, D89(3):033009, 2014.

\bibitem{Chatrchyan:2013gfi}
{Study of the Underlying Event at Forward Rapidity in pp Collisions at
  $\sqrt{s}$ = 0.9, 2.76, and 7 TeV}.
\newblock {\em JHEP}, 04:072, 2013.

\bibitem{Khachatryan:2015pea}
{Event generator tunes obtained from underlying event and multiparton
  scattering measurements}.
\newblock {\em Eur. Phys. J.}, C76(3):155, 2016.

\bibitem{Davidson:2010rw}
N.~Davidson, G.~Nanava, T.~Przedzinski, E.~Richter-Was, and Z.~Was.
\newblock {Universal Interface of TAUOLA Technical and Physics Documentation}.
\newblock {\em Comput. Phys. Commun.}, 183:821--843, 2012.

\bibitem{Agostinelli:2002hh}
S.~Agostinelli et~al.
\newblock {GEANT4: A Simulation toolkit}.
\newblock {\em Nucl. Instrum. Meth.}, A506:250--303, 2003.

\bibitem{Khachatryan:2016kdb}
{Jet energy scale and resolution in the CMS experiment in pp collisions at 8
  TeV}.
\newblock {\em JINST}, 12(02):P02014, 2017.

\bibitem{CMS:MCZee}
Dytoee\_m-20\_ct10\_tunez2star\_v2\_8tev-powheg-pythia6 in aodsim format for
  2012 collision data.
\newblock CERN Open Data Portal: \url{http://opendata.cern.ch}.
\newblock DOI:10.7483/OPENDATA.CMS.MXZU.YILG.

\bibitem{CMS:MCZmumu}
Dytomumu\_m-20\_ct10\_tunez2star\_v2\_8tev-powheg-pythia6 in aodsim format for
  2012 collision data.
\newblock CERN Open Data Portal: \url{http://opendata.cern.ch}.
\newblock DOI:10.7483/OPENDATA.CMS.QGC3.PTZ9.

\bibitem{CMS:MCZLL}
Simulated dataset
  dyjetstoll\_m-50\_tunez2star\_8tev-madgraph-tarball-tauola-taupolaroff in
  aodsim format for 2012 collision data.
\newblock CERN Open Data Portal: \url{http://opendata.cern.ch}.
\newblock DOI:10.7483/OPENDATA.CMS.DRSP.TO3O.

\bibitem{CMS:MCWPMunu}
Simulated dataset wplustomunu\_ct10\_8tev-powheg-pythia6 in aodsim format for
  2012 collision data.
\newblock CERN Open Data Portal: \url{http://opendata.cern.ch}.
\newblock DOI:10.7483/OPENDATA.CMS.I3N4.AVW3.

\bibitem{CMS:MCWMMunu}
Simulated dataset wminustomunu\_ct10\_8tev-powheg-pythia6 in aodsim format for
  2012 collision data.
\newblock CERN Open Data Portal: \url{http://opendata.cern.ch}.
\newblock DOI:10.7483/OPENDATA.CMS.F78O.K8QJ.

\bibitem{CMS:MCWPTau}
Simulated dataset wplustotaunu\_ct10\_8tev-powheg-pythia6-tauola in aodsim
  format for 2012 collision data.
\newblock CERN Open Data Portal: \url{http://opendata.cern.ch}.
\newblock DOI:10.7483/OPENDATA.CMS.E8GR.HZCR.

\bibitem{CMS:MCWMTau}
Simulated dataset wminustotaunu\_ct10\_8tev-powheg-pythia6-tauola in aodsim
  format for 2012 collision data.
\newblock CERN Open Data Portal: \url{http://opendata.cern.ch}.
\newblock DOI:10.7483/OPENDATA.CMS.67QN.1ZR0.

\bibitem{CMS:MCTTLep}
Simulated dataset ttjets\_fullleptmgdecays\_tunep11tev\_8tev-madgraph-tauola in
  aodsim format for 2012 collision data.
\newblock CERN Open Data Portal: \url{http://opendata.cern.ch}.
\newblock DOI:10.7483/OPENDATA.CMS.7RZ3.0BXP.

\bibitem{CMS:MCTTSemi}
Simulated dataset ttjets\_semileptmgdecays\_8tev-madgraph in aodsim format for
  2012 collision data.
\newblock CERN Open Data Portal: \url{http://opendata.cern.ch}.
\newblock DOI:10.7483/OPENDATA.CMS.FZCE.MBDW.

\bibitem{CMS:MCTTHad}
Simulated dataset ttjets\_hadronicmgdecays\_tunep11mpihi\_8tev-madgraph-tauola
  in aodsim format for 2012 collision data.
\newblock CERN Open Data Portal: \url{http://opendata.cern.ch}.
\newblock DOI:10.7483/OPENDATA.CMS.B91N.86OR.

\bibitem{CMS:MCWW}
Simulated dataset wwjetsto2l2nu\_tunez2star\_8tev-madgraph-tauola in aodsim
  format for 2012 collision data.
\newblock CERN Open Data Portal: \url{http://opendata.cern.ch}.
\newblock DOI:10.7483/OPENDATA.CMS.V2C6.O1P4.

\bibitem{CMS:MCWZ}
Simulated dataset wzjetsto3lnu\_8tev\_tunez2star\_madgraph\_tauola in aodsim
  format for 2012 collision data.
\newblock CERN Open Data Portal: \url{http://opendata.cern.ch}.
\newblock DOI:10.7483/OPENDATA.CMS.71R9.VLZA.

\bibitem{CMS:MCZZ}
Simulated dataset zzto4mu\_8tev-powheg-pythia6 in aodsim format for 2012
  collision data.
\newblock CERN Open Data Portal: \url{http://opendata.cern.ch}.
\newblock DOI:10.7483/OPENDATA.CMS.HJ1F.5U9R.

\bibitem{Khachatryan:2016bia}
{The CMS trigger system}.
\newblock {\em JINST}, 12(01):P01020, 2017.

\bibitem{Khachatryan:2016mqs}
{Measurement of the t-tbar production cross section in the e-mu channel in
  proton-proton collisions at sqrt(s) = 7 and 8 TeV}.
\newblock {\em JHEP}, 08:029, 2016.

\bibitem{Khachatryan:2014gga}
{Performance of the CMS missing transverse momentum reconstruction in pp data
  at $\sqrt{s}$ = 8 TeV}.
\newblock {\em JINST}, 10(02):P02006, 2015.

\bibitem{Aad:2019rou}
{Measurement of the cross-section and charge asymmetry of $W$ bosons produced
  in proton-proton collisions at $\sqrt{s}=8$ TeV with the ATLAS detector}.
\newblock 2019.

\bibitem{Khachatryan:2016crw}
{Measurements of differential production cross sections for a Z boson in
  association with jets in pp collisions at $ \sqrt{s}=8 $ TeV}.
\newblock {\em JHEP}, 04:022, 2017.

\bibitem{Khachatryan:2016poo}
{Measurement of the WZ production cross section in pp collisions at $\sqrt{s} =
  7$ and 8 $\,\text{TeV}$ and search for anomalous triple gauge couplings at
  $\sqrt{s} = 8\,\text{TeV} $}.
\newblock {\em Eur. Phys. J.}, C77(4):236, 2017.

\end{thebibliography}


\end{document}
\endinput